\documentclass[5p,twocolumn,10pt,times]{elsarticle}
\usepackage{amsmath}
\usepackage{hyperref}
\usepackage[switch]{lineno}
\usepackage{placeins}
\usepackage{float}
\graphicspath{{./}}
\usepackage{my}

\begin{document}
\baselineskip11pt

\begin{frontmatter}

\title{Outer Contour-driven Ruled Surface Generation for Linear Hot-wire Rough Machining}

\author[one]{Zheng Zhang} \ead{zheng1003@mail.ustc.edu.cn}
\author[zero]{Kang Wu\corref{cor}} \ead{kang910042009@gmail.com}
\author[zero]{Yi-Fei Li} \ead{lyf135266@mail.ustc.edu.cn}
\author[two]{Xu Liu} \ead{xuliutj@outlook.com} 
\author[two]{Xiang Wang} \ead{18310021@tongji.edu.cn} 
\author[zero]{Ligang Liu} \ead{lgliu@ustc.edu.cn} 
\author[zero]{Xiao-Ming Fu} \ead{fuxm@ustc.edu.cn}
\cortext[cor]{Corresponding authors}

\address[one]{School of Artificial Intelligence and Data Science, University of Science and Technology of China, Hefei, 230026, People's Republic of China}
\address[zero]{School of Mathematical Sciences, University of Science and Technology of China, Hefei, 230026, People's Republic of China}
\address[two]{College of Architecture and Urban Planning of Tongji University, Shanghai, 200092, People's Republic of China}



\begin{abstract}
We propose a novel method to generate a small set of ruled surfaces that do not collide with the input shape for linear hot-wire rough machining.
    Central to our technique is a new observation: the ruled surfaces constructed by vertical extrusion from planar smooth curves, which approach the input shape's outer contour lines while having no collisions, are capable of removing materials effectively during rough machining.
    %
    %
    %
    %
    Accordingly, we develop an iterative algorithm that alternates in each iteration between computing a viewpoint to determine an outer contour line and optimizing a smooth curve to approximate that contour line under the collision-free constraint.
    Specifically, a view selection approach based on genetic algorithm is used to optimize the viewpoint for removing materials as much as possible, and present an adaptive algorithm to find the constrained curves.
    The feasibility and practicability of our method are demonstrated through 10 physical examples.
    Compared with manual designs, our method obtains lower errors with the same number of cuts.  
\end{abstract}


\end{frontmatter}


\section{Introduction} \label{sec:intro}
\begin{figure}[t]
  \centering
  \begin{overpic}[width=0.99\linewidth]{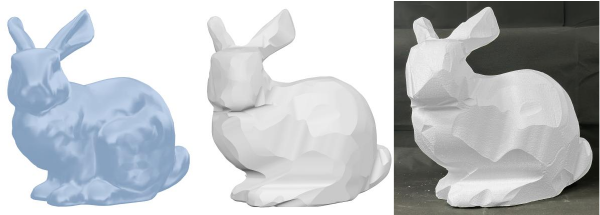}
    {
    }
  \end{overpic}
  \vspace{-3mm}
  \caption{
   We generate a small set of ruled surfaces to guide the cutting of material surrounding the input shape (left) for rough machining.
  The simulation result (middle) and physically fabricated model (right) are shown.
  }
  \label{fig:rough-machining}
\end{figure}

Rough machining is an initial phase in the subtractive manufacturing process.
In this step, the materials should be removed quickly to produce an initial shape that encloses the desired shape tightly for the subsequent fine machining process (see Fig.~\ref{fig:rough-machining}).
Compared to fine machining, rough machining does not impact the quality of the final products; however, it is still crucial throughout the manufacturing process because it directly affects the processing efficiency.

Common rough machining methods include milling \cite{yan2018multi, LIU2021360, LI202495}, wire cutting \cite{CHEN2018557, xu2023complex, zhang2025carving}, etc.
For wire cutting applications, a straight-wire configuration delivers more stable cutting velocity and better controllability.
%
The surface produced in this way is the ruled surface formed by the trajectory of a straight line running in space.
In general, these ruled surfaces should satisfy the following requirements.
First, each ruled surface should not collide with the input shape to ensure the completeness of the result.
Second, the number of ruled surfaces is small to accelerate the cutting process.
Third, when cutting along these ruled surfaces, as much material as possible should be removed to reduce the difficulty of fine machining.

However, automatically constructing such ruled surfaces is difficult, and the reasons are two-fold.
First, reducing the number of ruled surfaces and increasing material removal is inconsistent, making it difficult to achieve a favorable trade-off.
Second, the collision-free constraint is non-linear and non-convex, thereby increasing the difficulty in determining the position and shape of the ruled surface.

%
%
%

\cite{van2019accessibility} conduct preliminary research on this problem.
They can guarantee collision-free paths, but need users to input cutting paths, and cannot handle complex shapes. 
\cite{duenser2020robocut} cut materials with curved hot wires.
Their key contributions are computing the relationship between the hot wire's shape and the robot's motions and approximating an input shape with surfaces.
However, they use one-sided quadratic penalty functions to avoid collisions, which do not strictly guarantee that there will be no collisions.
In addition, although their method has an automatic cut initialization module, it only samples some cutting paths and picks among them.
The lack of optimization for the cut initialization lowers their efficiency. 
Generally, there is no effective automatic algorithm for cutting path planning in rough machining.

\begin{figure}[t]
  \centering
  \begin{overpic}[width=0.97\linewidth]{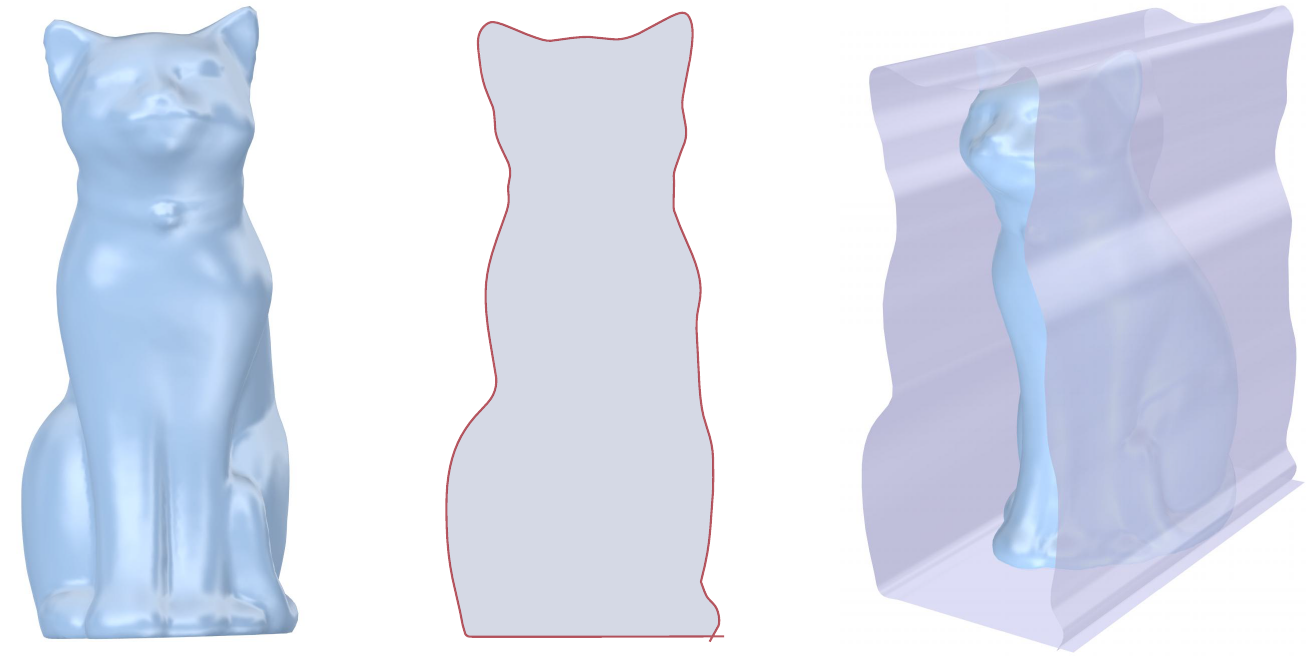}
    {
 \put(10,-3){\small \textbf{(a)}}
 \put(43,-3){\small \textbf{(b)}}
 \put(80,-3){\small \textbf{(c)}}
    }
  \end{overpic}
  \vspace{0mm}
  \caption{
  (a) A viewpoint of the cat model. (b) A smooth curve (red) that approximates the outer contour line of this viewpoint without collisions. (c) The ruled surface (light purple) extruded from the smooth curve.
  }
  \label{fig:outer-contour-line}
\end{figure}

In this paper, we propose a novel method to generate a small set of ruled surfaces, achieving a favorable trade-off between the requirements.
The algorithm relies on a key observation.
We observe that using the outer contour line under a specific viewpoint to generate a ruled surface as the following two steps can remove materials effectively (see Fig.~\ref{fig:outer-contour-line}).
First, we use a smooth curve to approach and enclose the outer contour line under the collision-free constraint as the base curve of the ruled surface.
Second, we extrude this curve along the vertical direction of the camera plane to form the ruled surface.
%
The ruled surface constructed in this way has two advantages in rough machining: (1) the ruled surface does not collide with the input shape and is close to the input shape, and (2) the ruled surface is sufficiently long as the outer contour line is long, enabling the robot arm to remove as much material as possible when cutting along the ruled surface.

Accordingly, 
We develop a greedy algorithm to generate ruled surfaces iteratively.
In each iteration, our algorithm has the following three steps: (1) viewpoint selection, (2) smooth curve approximation, and (3) ruled surface extrusion to remove material outside.
Specifically, we introduce a viewpoint selection for removing materials as much as possible based on a genetic algorithm.
After obtaining the outer contour line using the selected viewpoint, we present an adaptive technique to optimize a desired smooth curve.
We evaluate our method using a variety of examples to demonstrate its feasibility and effectiveness. 
We also showcase its usability and practicality by fabricating 10 models using the linear hot-wire cutting technique.
%
Compared with manual design, our method achieves a smaller approximation error with an equal number of cuts.

\section{Related work} \label{sec:related}

\paragraph{Ruled surface-driven machining}
In practical machining processes, many shapes of cutting tools, such as straight lines, cylinders, and so on, generate ruled surfaces when cutting materials.
Thus, many researchers studied how to carve shapes with ruled surfaces.
~\cite{lin2000ruled} developed a real-time five-axis interpolator for ruled surface machining, which continuously aligns tool orientation with ruling lines, achieving higher accuracy and productivity compared to conventional offline methods.
~\cite{koc2002adaptive} proposed an adaptive ruled layer approximation method for stereolithography(STL) models, where adjacent contours are connected by ruling lines to form ruled surfaces. 
A surface error analysis identifies maximum error at each rapid prototyping layer, which then guides the generation of adaptive ruled layers for STL CAD models.
~\cite{sprott2008cylindrical} developed a method for cutting path planning in flank or cylindrical milling.
Their method computes the position of the cutting tool by utilizing the geometric properties of the ruled surface, which greatly simplifies the error calculation.
~\cite{gong2005improved} reformulated this problem to optimize the deviation between the offset surface of the target surface and the surface formed by the tool axis trajectory.
They proposed a points offset strategy to estimate the offset surface and used a simple least square approximation method to reduce errors further. However, their algorithm lacks an automated initialization of guiding curves.
The case with a conical tool was studied by~\cite{li2006flank}.
The cutting tool operates tangentially to two guiding rails that can be located anywhere on the ruled surface.
Given the initial position of the cutting tool, their method reduces surface error by adjusting the guiding rails toward the point of greatest deviation along the ruling line and repositioning the tool in the feed direction.
Recently, two hot-wire cutting methods were proposed by~\cite{hua2018wire} and~\cite{van2019accessibility}, respectively.
The method of Hua and Jia produces double-sided minimal surfaces with a wire-cut machine.
It cuts along the principal curvature lines of the surface to remove materials from both sides.
van Sosin et al. proposed a method to find collision-free tangential cutting directions.
Along these directions, the cutting paths are globally accessible during the machining process.
~\cite{CHU2020171} developed a tool path generation method for minimizing geometrical errors on finished ruled surfaces while preserving high-order continuity in the cutter motion.
%
~\cite{zhou2021digital} presents a digital-twin–driven framework: first, it transforms free-form blades into near-developable ruled surfaces; second, it generates NURBS-based flank-milling toolpaths; finally, it applies reinforcement learning to balance machining efficiency and aerodynamic performance.

\paragraph{Next best view}
The next best view problem is a classical one in computer graphics.
%
%
%
\cite{yamauchi1997frontier} uses voxel grids to represent the object to be observed.
He denotes the border of free and unknown voxels as frontiers and utilizes this information to calculate the next best view.
Other volumetric-based methods~\cite{vasquez2018tree, vasquez2014view, vasquez2017view, vasquez2009view, vasquez2014volumetric, potthast2014probabilistic, delmerico2018comparison} apply a ray traversal algorithm to estimate the unknown voxels so that they can collect the whole information of the view.
Then, the NBV metrics are computed based on the information and used to select the next viewpoint from among all candidate views.
Surface-based methods~\cite{wu2014quality, maver1993occlusions, yuan1995mechanism, pito1999solution, chen2005vision, zhou2009novel, kriegel2011surface} represent the input scene as a mesh, point cloud, signed distance, and so on.
These methods utilize the geometry information, such as contour or curvature tendency, to guide NBV selection.
The above methods either use a ray traversal algorithm to compute the information used for selecting viewpoints or utilize the geometry information, such as curvature tendency, to guide NBV selection, which limits the efficiency.
%
%
%
In recent years, with the development of neural networks, learning-based methods~\cite{mendoza2020supervised, wu20153d, zeng2020pc, liu2024dg, xiao2024next, li2024boundary, jin2023neu} 
have been proposed to solve the NBV planning problem.
%
%
These methods cover all input model representations and perform better than previous works, but they need to spend time training their network.
The neural implicit field is also used in the NBV planning problem.
The methods~\cite{lee2022uncertainty, ran2023neurar, pan2022activenerf, yan2023active} exploit the differentiability of the implicit field to construct a differentiable objective function.
Thus, the gradient-based optimization algorithm can be used to solve the problems efficiently.
However, converting the input model into an implicit field requires additional time.
%


\paragraph{B-Spline fitting}
Many methods have been proposed to address the curve fitting problem with B-spline curves.
A traditional technique assigns a parameter value to each fitted data point~\cite{park2001choosing, vassilev1996fair, wang1997energy}.
This method needs to choose a suitable data parametrization as the fitting results vary with changes in parametrization~\cite{dierckx1995curve}.
Another technique for solving the problem is to use the sum of the squares of the distances between the data points and the curve.
According to the distance used in the objective function, the algorithms can be divided into three parts: point distance~\cite{zheng2012fast, javidrad2012accelerated, lin2019certified, jover2022coupled, wang2006fitting, pottmann2002approximation}, tangent distance~\cite{liu2008revisit, blake2012active}, and squared distance~\cite{wang2006fitting}.
Different optimization techniques are applied for curve fitting, such as the diagonal approximation BFGS method~\cite{ebrahimi2019b}, the variable projection method~\cite{bergstrom2012fitting}, the conjugate gradient method~\cite{xie2001automatic}, and some discrete optimization algorithms~\cite{irshad2016outline, galvez2013firefly, javidrad2012accelerated, hasegawa2014bezier}.
Active contour models are widely used in image segmentation tasks where curves are deformed to align with image features such as lines and edges~\cite{zhang2020active, monemian2020analysis, sasmal2021detection, yang2022active, kass1988snakes}.
Moreover, these models are also used in curve fitting problems~\cite{pottmann2002approximation, bayer2005laplace, wang2006fitting}.
We consistently ensure collision-free constraints from the outset and use a progressive strategy to minimize fitting errors through adaptive parameterizations and incremental addition of control points.

\section{Method} \label{sec:method}
\subsection{Overview}


\paragraph{Inputs and goals}
Our inputs include a shape represented as a triangular mesh $\mathcal{M}$ 
and a material box $\mathcal{B}$ that encloses $\mathcal{M}$.  
%
%
%
The goal is to automatically generate a small set of ruled surfaces $\mathcal{S}=\{\ms_i\}$ satisfying the following requirements:
\begin{itemize}
\item \emph{Cutting validity:} the ruled surfaces $\mathcal{S}$ should not collide with the input shape $\mathcal{M}$.
    \item \emph{Cutting quality:} cutting along the ruled surfaces can remove as many materials as possible.
      \item \emph{Cutting efficiency:} the number of the ruled surfaces is small.  
\end{itemize}
The cutting validity requirement is a hard constraint.
Without loss of generality, we assume that the material $\mathcal{B}$ is a box with a side length of 1, and its center is at the origin.
We scale the input shape $\mathcal{M}$ into the box $\mathcal{B}$. 

\IncMargin{0.5em}
\begin{algorithm}[t]
	\caption{Workflow}\label{alg:workflow-code}
	\newcommand\mycommfont[1]{\footnotesize\ttfamily\textcolor{blue}{#1}}
	\SetCommentSty{mycommfont}
	\SetKwInOut{AlgoInput}{Input}
	\SetKwInOut{AlgoOutput}{Output}

	\SetKwFunction{OV}{ViewpointSelection}
    \SetKwFunction{OOC}{ObtainOuterContour}
        \SetKwFunction{SCA}{2DSmoothCurveApproximation}
        \SetKwFunction{GRS}{ExtrudeRuledSurface}
        \SetKwFunction{CARS}{CutAlongRuledSurface}
\SetKwFunction{PIS}{PushIntoRuledSurfaceSet}
	\AlgoInput{the input shape $\mM$, the material box $\mB$}
	\AlgoOutput{the set of ruled surfaces $\mS$}
        $k \leftarrow 1;\mB_k \leftarrow \mB;\mS \leftarrow \emptyset$\;
	\While{not terminated}{
        \hspace*{-0.2em}$C_k \leftarrow \OV(\mM, \mB_k)$;\tcc{Section~\ref{sec:viewpoint-opt}}\
        \hspace*{-0.3em}$\mg \leftarrow \OOC(\mM, C_k)$\;
        \hspace*{0em}\mbox{$\mc \leftarrow \SCA(\mg)$;\tcc{Section~\ref{sec:2d-fitting}}}\
        \hspace*{-0.5em}\mbox{$\mathbf{s} \leftarrow \GRS(\mc)$;\tcc{Section~\ref{sec:extrusion}}}\
        \hspace*{-0.7em}$\mB_{k+1} \leftarrow \CARS(\mathbf{s},\mB_k)$\;
        $\PIS(\mathbf{s}, \mS)$\;
        $k \leftarrow k+1$\;
	}
\end{algorithm}
\DecMargin{0.5em}

\paragraph{Pipeline}
Our method first utilizes the genetic algorithm to find the proper viewpoint (Section~\ref{sec:viewpoint-opt}).
%
%
%
Since using smooth surfaces as cutting paths can ensure the hot wire moves uniformly to avoid material over-burning, we use B-spline surfaces to represent the ruled surfaces.
Then, we propose a 2D B-spline fitting algorithm to approximate the outer contour line under the proper viewpoint (Section~\ref{sec:2d-fitting}) and extrude it to generate the ruled surface (Section~\ref{sec:extrusion}).
Then, we cut along the ruled surface to remove the material.
%
These steps are iteratively performed until the volume difference between the remaining material and the input shape is under a specific threshold $\alpha$ or the iteration number exceeds the specified maximum number $N_\text{iter}$.
Alg.~\ref{alg:workflow-code} shows the pseudo-code.

\subsection{Viewpoint selection}\label{sec:viewpoint-opt}

\subsubsection{Problem}\label{sec:fitnessProblem}
\paragraph{Inputs and goals}
Given the input shape $\mathcal{M}$ and the remaining material $\mathcal{B}_i$ after the $(i-1)$-th cut, our goal is to determine a viewpoint such that the ruled surface generated from the outer contour line of $\mathcal{M}$ under this viewpoint allows the subsequent cut to remove as much material as possible, thereby reducing the difficulty of later iterations and fine machining. 

\paragraph{Formulation}
This task naturally falls into the framework of a next-best-view (NBV) problem, where the algorithm iteratively selects a new viewpoint to achieve maximal material reduction. The optimization problem is:
\begin{equation}\label{equ:viewpoint-obj}
   \max_{\mathbf{C}_i} \quad E_\text{material} = \lVert(\mathcal{I}^a_{\mathcal{B}_i} - \mathcal{I}^a_\mathcal{M}) \odot \mathcal{I}^v_{\mathcal{B}_i}\rVert_F^2,
\end{equation}
where $\mathbf{C}_i$ is the camera parameter in $i$-th iteration, $\mathcal{I}^a_{\mathcal{B}_i}$ and $\mathcal{I}^a_{\mathcal{M}}$ are the 
\emph{area images} of $\mathcal{B}_i$ and $\mathcal{M}$ under $\mathbf{C}_i$, respectively, $\mathcal{I}^v_{\mathcal{B}_i}$ is the \emph{volume image} of $\mathcal{B}_i$ rendered with $\mathbf{C}_i$, and $\odot$ is the Hadamard product of two matrices.
The pixel value of an area image of an object is 1 if the pixel is in the object's rendering image; otherwise, it is 0.
The pixel value of a volume image represents the volume of the part of the object whose projection is located in this pixel.

\paragraph{Our objective function}
Using volume rendering technique to obtain $\mathcal{I}^v_{\mathcal{B}_i}$ is 
inefficient as the shape of $\mathcal{B}_i$ can be complex.
Thus, we use the projected area of $\mathcal{B}_i$ to approximate its volume.
In this case, our objective function is reformulated as follows:
\begin{equation}\label{equ:our-viewpoint-obj}
   E^\text{ours}_\text{material} = \lVert\mathcal{I}^a_{\mathcal{B}_i} - \mathcal{I}^a_\mathcal{M}\rVert_F^2.
\end{equation}
This objective function is an approximation of $E_\text{material}$. 
When the area difference is small enough, the surface of the remaining materials is close enough to the input shape.

\subsubsection{Method overview}\label{sec:fitnessMethod}
To efficiently find the desired viewpoint in the $i$th iteration, we adopt the genetic algorithm to solve this NBV problem.
Initially, our genetic algorithm selects a subset of candidate viewpoints to form an initial population, where each individual in the population corresponds to a viewpoint (Section~\ref{sec:Initialization}). 
In each subsequent iteration, we update the population by performing three genetic operations in sequence (Section~\ref{sec:Operations}): crossover, mutation, and selection, while keeping the population size constant. 
After the iterations terminate (Section~\ref{sec:other-details}), the individual with the highest objective function value, called the fitness value, is chosen as the best viewpoint $\mp$.
The best viewpoint is subsequently used to guide ruled surface generation for material removal. 
We show the pseudocode of the genetic algorithm in Alg.~\ref{alg:genetic}.

\subsubsection{Rendering settings}
We adopt Pytorch3D~\cite{ravi2020pytorch3d} to render the area images $\mathcal{I}^a_{\mathcal{B}_i}$ and $\mathcal{I}^a_{\mathcal{M}}$.
The shader is \textit{HardPhongShader}. 
Our parameters in the shader are the default.
We use the orthogonal projection to render $\mathcal{I}^a_{\mathcal{B}_i}$ and $\mathcal{I}^a_{\mathcal{M}}$.
The image resolution is $256\times256$.
The projected area approaches 0 when the camera is far from the rendered object.
To avoid this, we limit the camera's position to a sphere surface with a radius of $r$ and a center located at the origin.
So, the camera position $(x,y,z)$ is:
\begin{equation}\label{equ:camera-pos}
\left\{
\begin{aligned}
x & =  r \cos\phi  \cos \theta  \\
y & =  r \cos\phi  \sin \theta  \\
z & =   r \sin\phi 
\end{aligned}
\right.
,
\end{equation}
Where $\phi$ and $\theta$ are the elevation and azimuth, respectively. 
In our experiments, we set $r=2$. 
In this case, the camera parameter $\mathbf{C}_i$ is the elevation and azimuth $(\phi_i, \theta_i)$

\IncMargin{0.5em}
\begin{algorithm}[t]
	\caption{Genetic viewpoint selection framework}\label{alg:genetic}
	\newcommand\mycommfont[1]{\footnotesize\ttfamily\textcolor{blue}{#1}}
	\SetCommentSty{mycommfont}
	\SetKwInOut{AlgoInput}{Input}
	\SetKwInOut{AlgoOutput}{Output}

	\SetKwFunction{OV}{InitializePopulation}
    \SetKwFunction{OOC}{Selection}
        \SetKwFunction{SCA}{Crossover}
        \SetKwFunction{GRS}{Mutation}
        \SetKwFunction{FBC}{FindBestChromosome}
	\AlgoInput{the input shape $\mM$, the material box $\mB$}
	\AlgoOutput{the best viewpoint $\mp$}
        $k \leftarrow 1$;$\mA_k \leftarrow \OV(\mM, \mB_k)$\;\tcc{Section~\ref{sec:Initialization}}\
	\While{not terminated \tcc{Section~\ref{sec:Operations}}}
    {
        $\mA_k \leftarrow \OOC(\mA_k)$\;
        $\mA_k \leftarrow \SCA(\mA_k)$\;
        $\mA_k \leftarrow \GRS(\mA_k)$\;
        $k \leftarrow k+1$\;
	}
        $\mp \leftarrow \FBC(\mA_k)$;\tcc{Section~\ref{sec:Operations}}\
\end{algorithm}
\DecMargin{0.5em}

\subsubsection{Initialization}\label{sec:Initialization}
To improve the search efficiency of our algorithm, we do not solve our viewpoint selection problem in the continuous hemisphere parameter space.
Instead, we use the Fibonacci grid sampling method to pre-sample this parameter space to form a discrete set $\mathcal{P} = \{\mathbf{p}_i = (\phi_i,\theta_i)\}_{i=1}^{N_\text{sample}}$ and run our genetic algorithm on this set to find the best viewpoint.
Constraining our method to search on these discrete candidates enables our genetic algorithm to converge quickly. 

After obtaining our discrete searching set $\mathcal{P}$, we perform the farthest-point sampling method on $\mathcal{P}$ to select $N_{\text{pop}}$ viewpoints to form the initial population $\mA_0$.
By sampling in this way, we ensure the initial viewpoints are spread as evenly as possible over the hemisphere, improving initial diversity and reducing the likelihood of redundant evaluations in the early generations. 
In our experiment, we set $N_\text{sample} =5000$ and $N_{\text{pop}} = 30$.
\subsubsection{Operations}\label{sec:Operations}
Our genetic algorithm consists of three operations: crossover, mutation, and selection.
Assuming in $i\text{th}$ iteration, the population is $\mathcal{A}_i$
The crossover operation is used to create a new individual (child) from two parents, which means we calculate a new candidate viewpoint based on two viewpoints selected in $\mathcal{A}_i$.
The mutation operation is applied to the children generated by the crossover.
This operation aims to change the children's value to help our method avoid getting stuck in local optima.
The last operation, selection, is performed to generate a new population $\mathcal{A}_{i+1}$ based on $\mathcal{A}_{i+1}$ and the children we generate after the crossover and mutation operations.

\paragraph{Crossover}
To effectively explore the discrete space of candidate viewpoints while preserving and recombining high-quality solutions, we apply a crossover operator to generate new child individuals from pairs of parents. 

In this operation, parents are selected using a roulette-wheel selection method, in which each individual is chosen with a probability proportional to its fitness energy $E^\text{ours}_\text{material}$.
The selection probability for the $i$‑th chromosome is computed as:
$$
p_i^s = \frac{E^\text{ours}_\text{material}(\textbf{p}_i)}{\sum_{j=1}^{N_\text{pop}} E^\text{ours}_\text{material}(\textbf{p}_j)},
$$
where $E^\text{ours}_\text{material}(\textbf{p}_i)$ denotes the fitness of individual $\textbf{p}_i$, and $N_\text{pop}$ is the population size. 
%
%
This approach increases the likelihood that high-fitness individuals will be selected for reproduction, while still allowing lower-fitness individuals to contribute occasionally, thereby preserving genetic diversity within the population. 
%
%

We perform $N_{\text{pop}}$ selections on $\mathcal{A}_i$ to form the set of parents, selecting an individual from $\mathcal{A}_i$ each time. 
The probability of each individual being selected is $p_i^s$.
Then we simply pair the selected parents up to generate children.
Once the same pair of parents appears, we delete some of them to keep only one pair and then perform the selection operation again and again until all parent pairs are different.
The children are generated by the parent pairs using a sequence of three probabilistic crossover strategies: uniform crossover, geometric midpoint crossover, and neighborhood crossover.
These three strategies are applied independently to each child.

The uniform crossover is applied with probability $p_\text{uni}$. 
For each child $\mathbf{p}^c_i$, let $\mathbf{p}^f_i$ and $\mathbf{p}^m_i$ denote the corresponding parents. 
Then the child is set as:
$$
\mathbf{p}^c_i = 
\begin{cases}
\mathbf{p}^f_i, & \text{with probability } p^f_u, \\
\mathbf{p}^m_i, & \text{otherwise}.
\end{cases}
$$
This mechanism enables large, non-directional jumps in the discrete viewpoint space and promotes population diversity.

The geometric midpoint crossover is performed with probability $p_\text{geo}$.
%
%
Given two parents $\mathbf{p}^f_i$ and $\mathbf{p}^m_i$, the child $\textbf{p}^m_i$ generated by the geometric midpoint crossover is computed by a spherical linear interpolation:
$$
\mathbf{p}_i^m = \frac{\sin((1-t)\omega)\,\mathbf{p}^f_i + \sin(t\omega)\,\mathbf{p}_i^m}{\sin(\omega)},
\quad \omega = \arccos(\mathbf{p}_i^f \cdot \mathbf{p}_i^m).
$$
$\mathbf{p}_i^m$ is then projected back to the nearest viewpoint in the discrete candidate set $\mathcal{A}_i$. 
This strategy allows for structured interpolation between two viewpoints, producing child that lie between their parents in angular space. In our experiment, we set $t = 0.5$.

The neighborhood crossover is applied with probability $p_\text{nei}$. 
In this operation, our algorithm uniformly randomly selects one individual of the two parents' $k$-nearest neighbors ($k=10$ in our experiments) to replace the child.
%
%
This enables small-scale, directionally coherent modifications that preserve local structure and promote fine-tuning around high-fitness regions.

By performing the above three strategies with a certain probability, our crossover operation effectively balances global exploration and local refinement.
This allows the algorithm to search widely across the hemisphere while performing precise optimization within high-fitness regions, thereby accelerating convergence and improving overall solution quality. 
In our experiments, we set $p_\text{uni} = 0.30$, $p_\text{geo} = 0.40$ and $p_\text{nei} = 0.30$.

\paragraph{Mutation}
Mutation is applied exclusively to the children generated by the crossover stage. 
Each child has a probability of undergoing one of two mutations - the global reset mutation and the local neighborhood mutation.

A global reset mutation occurs with probability $p_\text{glo}$: the child is mutated by selecting uniformly at random one viewpoint from the candidate set $\mathcal{C}$, enabling a large-scale jump across the entire discrete space. 
A local neighborhood mutation is employed with probability $p_\text{loc}$: the child is replaced by one of its $k$-nearest neighbors ($k=10$ in our experiments), affecting a local perturbation. 
If no mutation is triggered, the child remains unchanged, preserving the structural information inherited through crossover.
%

Mutation randomly perturbs an individual’s genes, enabling the algorithm to escape local optima. 
This prevents premature convergence and increases the likelihood of finding the global optimum. 
In our experiments, we set $p_\text{glo} = 0.05$ and $p_\text{loc} = 0.10$.

\paragraph{Selection}
To guide the evolution toward high‑fitness viewpoints, we adopt an elitist selection to form the new population for the next iteration.
Specifically, after performing crossover and mutation operations on $\mathcal{A}_i$ to generate new children of the population, we select the top $N_\text{pop}$ individuals with the highest fitness values $E_\text{fit}$ from $\mathcal{A}_i$ and the generated children to construct the population of the next generation $\mathcal{A}_{i+1}$.
The selection operation does not involve crossover or mutation.
This ensures that the highest-quality solutions are retained across generations, leading to a steady increase in the average fitness of the population as evolution progresses.

\subsubsection{Termination}\label{sec:other-details}
We terminate our genetic algorithm if the iteration reaches the maximum iteration $N_\text{max} =15$ or the population remains unchanged in $N_\text{converge}= 5$ successive iterations. 
The viewpoint with the highest fitness value $E^\text{ours}_\text{material}$ is the output of our genetic algorithm.

\subsection{Ruled surface generation}
\subsubsection{2D approximation}\label{sec:2d-fitting}
\paragraph{Problem}
After getting the optimal viewpoint, the next step is to create a ruled surface using the outer contour line of this viewpoint.
According to our observation, the ruled surface can be generated by extruding from a smooth curve that approaches the outer contour line without collisions.
%
In the actual manufacturing process, the speed of the cutter changes rapidly when the cutter changes direction at the turning points of polyline paths. 
Uneven cutter speed can lead to inconsistent heating times of the material, which can have a negative impact on manufacturing accuracy.
So we use a B-spline curve to approximate the outer contour line to serve as the cut path.
%
%
This problem can be formulated as a 2D B-spline curve approximation problem.
%
Given the outer contour line represented as a polygon $\mg$, the generated B-spline curve $\mathbf{c}(\mathbf{P}, \mathbf{U})$, where $\mathbf{P}$ is the set of control points and $\mathbf{U}$ is the set of knot vector values, should satisfy the following two requirements: (1) $\mathbf{c}$ encloses $\mg$ with a small approximation error; 
(2) there is no collision between $\mg$ and $\mathbf{c}$.
Without loss of generality, we set the center of $\mg$ as the origin for the approximation process.

%
%
%
%
%

\paragraph{Method overview}
To generate a constraint-satisfied result, we first generate a collision-free initial B-spline curve $\mathbf{c}$ and then keep this constraint satisfied when reducing the approximation error (Alg.~\ref{alg:pseudo-code}). 
%
We place the initial control points of $\mathbf{c}(\mathbf{P}, \mathbf{U})$ on a circle centered on the center of $\mg$.
As long as the circle radius is large enough, the collision-free and enclosing constraints can be easily satisfied.
%
%
Then, we progressively optimize $\mathbf{c}$, including control point position optimization, adaptive parameterizations, and incremental addition of control points, to reduce the error to a low level.

\paragraph{Objective function}
To generate a smooth B-spline curve approaching $\mg$ under the collision-free constraint, we construct our objective function based on the formulation of  ~\cite{wang2006fitting}.
It consists of an approximation error term and a smoothness term as follows:

\begin{equation}\label{equ:2D-objective}
   E = E_\text{error} + w E_\text{smooth},
\end{equation}
where
\begin{equation}\label{equ:2D-approximation-error}
\begin{aligned}
   &E_\text{error} = \sum_{\mathbf{v}_i \in S_g} \delta_{i}(E_{sd}-\epsilon),\\
    &E_{sd} =
    \raisebox{0.5ex}{\scalebox{0.8}{$
    \begin{cases}
   \frac{d}{d-\rho}[ (\mathbf{c}(u_i^c) - \mathbf{v}_i)^T T_k ]^2+[ (\mathbf{c}(u_i^c) - \mathbf{v}_i)^T N_k ]^2,& \text{if } d < 0,\\
    [ (\mathbf{c}(u_i^c) - \mathbf{v}_i)^T N_k ]^2,& \text{if } 0 \leq d < \rho,
    \end{cases}
    $}}\\
   &E_\text{smooth} = \sum_{\mathbf{c}(u_i) \in S_c}\eta_0\|\mathbf{c}'(u_i)\|_2^2+\eta_1\|\mathbf{c}''(u_i)\|_2^2,
\end{aligned}
\end{equation}

Here, $S_g$ and $S_c$ are the sets of uniformly sampling points in $\mg$ and $\mathbf{c}$ respectively (1000 points each in $\mg$ and $\mathbf{c}$), and $\mathbf{c}(u_i^c)$ is the corresponding point of $\mathbf{v}_i$.
%

In $E_\text{error}$, $T_k$ and $N_k$ are the unit tangent vector and the unit normal vector of the current fitting curve $\mathbf{c}(u)$ at the point $\mathbf{c}(u_i^c)$.
These two vector directions serve as the axes of the local Frenet frame of $\mathbf{c}(u)$ at $\mathbf{c}(u_i^c)$.
%
%
$\rho$ > 0 is the curvature radius of $\mathbf{c}(u)$ at $\mathbf{c}(u_i^c)$. 
We orientate the curve normal such that \( K = (0, \rho)^{T} \) is the curvature center of $\mathbf{c}(u)$ at $\mathbf{c}(u_i^c)$. 
Let \( d \) be the signed distance from  $\mathbf{v}_i$ to $\mathbf{c}(u_i^c)$ as follows:
%
\begin{equation}
d=
    \begin{cases}
   -\left\| \mathbf{c}(u_i^c) - \mathbf{v}_i \right\|,& \text{if}\  \mathbf{v}_i\ \text{and  K  are on opposite sides of}\ \mathbf{c}(u),\\
    \left\| \mathbf{c}(u_i^c) - \mathbf{v}_i \right\|,& \text{if}\  \mathbf{v}_i\ \text{and  K  are on the same side of}\ \mathbf{c}(u).
    \end{cases}
\end{equation}
We note that there is always \( d < \rho \) when \( d > 0 \), for otherwise \( \mathbf{c}(u_i) \) cannot be the closest point on the curve $\mathbf{c}(u)$ to $\mathbf{v}_i$.

$\delta_i$ is a function defined as follows:
\begin{equation}
\delta_i=
    \begin{cases}
   0,& \text{if}\ \ \overline{\mathbf{c}(u_i^c)\mathbf{v}_i} \cap \mathbf{g} \neq \emptyset,\\
    1,& \text{otherwise}.
    \end{cases}
\end{equation}
We use $\delta_i$ to eliminate the effect of the unsuitable corresponding points that cause the collisions between $\mathbf{c}$ and $\mathbf{g}$. 
$\epsilon$ is a positive constant that gradually decreases to 0 as the number of iterations increases.
It is introduced to control the convergence speed to avoid parts of $\mathbf{c}$ too close to $\mathbf{g}$ in advance so that their neighbor parts cannot continue moving.
In practice, the initial $\epsilon$ is set to $1\%d_\text{bb}$ (which denotes the length of the diagonal of the input shape's bounding box), and we multiply it by 0.8 in each iteration to decrease it.
We set $w=1$, $\eta_0=0$ and $\eta_1=1$in our experiments.
The collision-free constraint is not included in the objective function, and we use explicit checks to prevent collisions.

\begin{figure}[t]
  \centering
  \begin{overpic}[width=1.0\linewidth]{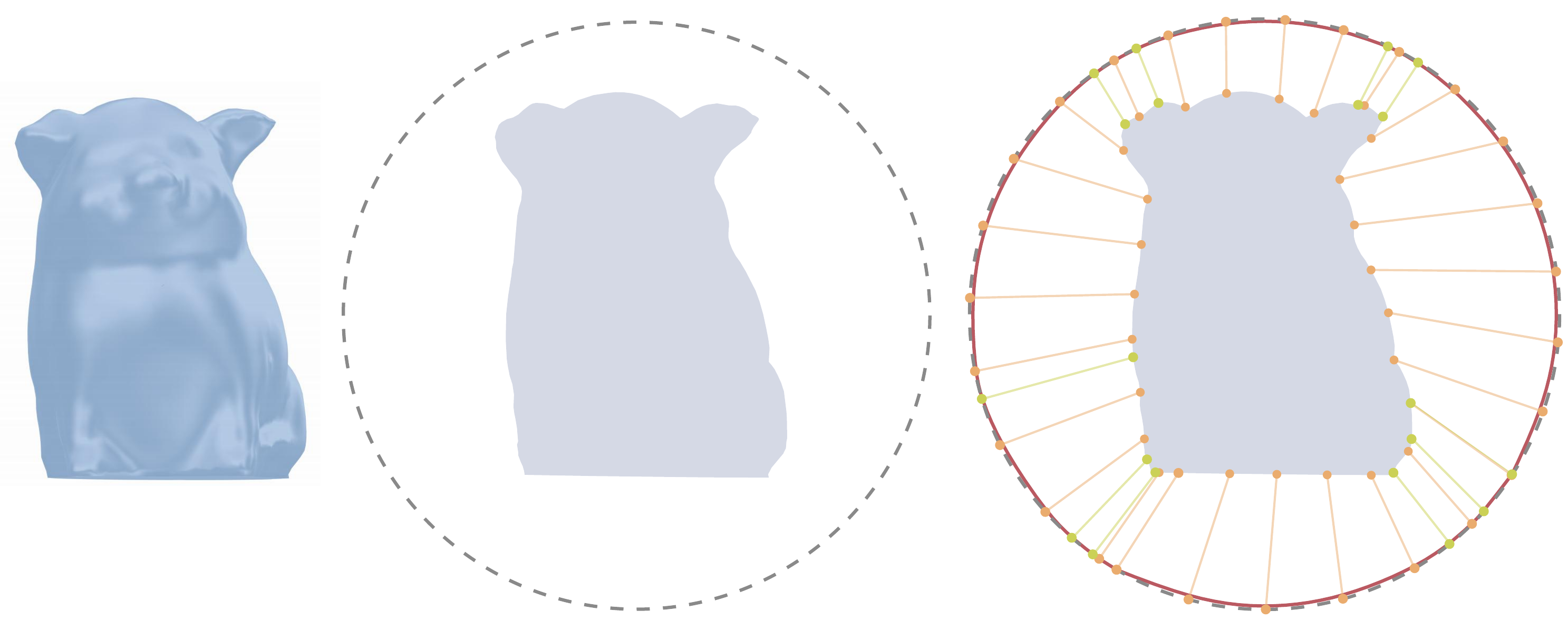}
    {
     \put(10,-3.9){\small \textbf{(a)}}
      \put(39,-3.9){\small \textbf{(b)}}
 \put(80,-3.9){\small \textbf{(c)}}
    }
  \end{overpic}
  \vspace{-3mm}
  \caption{
  Initializing control points. Given a viewpoint (a), we place initial control points on a circle outside the outer contour of this viewpoint (b).
  We color $\{\mathbf{p}_i^h\}$ and $\{\mathbf{v}_i^h\}$ in green, and $\{\mathbf{p}_i^s\}$ and $\{\mathbf{v}_i^s\}$ in orange (c).
  }
  \label{fig:init-control-points}
\end{figure}

\paragraph{Initialization}
The positions of the initial control points are constrained on a circle centered at the origin, denoted as $\mathbf{c}_r$ (see Fig.~\ref{fig:init-control-points} (b)).
We use the convex hull $\mh = \{\mathbf{v}_i^h\}$, where $\{\mathbf{v}_i^h\}$ is the subset of vertices of $\mg$, to determine the distribution of the control points.
For each $\mathbf{v}_i^h$, we compute its closest point $\mathbf{p}_i^h$ on $\mathbf{C}_r$ as an initial control point. 
The knot vector value $u_i^h$ of $\mathbf{p}_i^h$ is calculated as the ratio between the boundary length from $\mathbf{v}_i^h$ to $\mathbf{v}_0^h$ and the total boundary length.
%
%
For any point $\mathbf{p}_i$ in the arc from $\mathbf{p}_i^h$ to $\mathbf{p}_{i+1}^h$, we set its corresponding knot vector value $u_i = u_i^h + \frac{L(\mathbf{p}_i^h, \mathbf{p}_{i})}{L(\mathbf{p}_i^h, \mathbf{p}_{i+1}^h)} (u_{i+1}^h - u_i^h)$, where $L(\mx, \my)$ is the arc length between $\mx$ and $\my$. 
Then, we uniformly sample $N_\text{init}-N_h$ values in $[0, 1]$ as the knot vector values ($N_\text{init} = 40$ in our experiments, and $N_h$ is the number of vertices in $\{\mathbf{p}_i^h\}$) and compute the corresponding points $\{\mathbf{p}_i^s\}$ in $\mathbf{c}_r$ as the rest initial control points (see Fig.~\ref{fig:init-control-points} (c)).
The circle radius is initialized as 1.5 times the maximum distance from $\mg$ to the origin.
If the collision-free and enclosing constraints are not satisfied, we gradually enlarge the radius and run the above steps until the constraints are met.

\begin{figure}[t]
  \centering
  \begin{overpic}[width=1.0\linewidth]{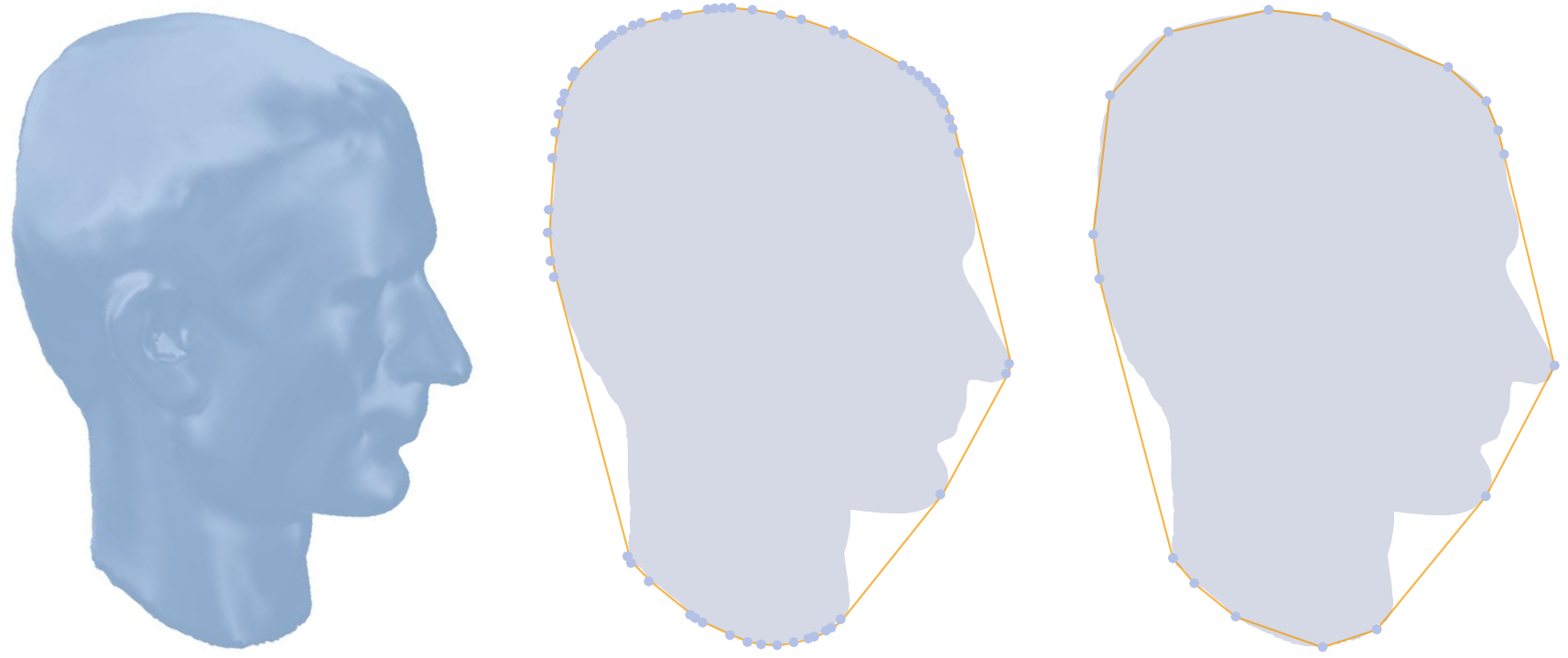}
    {
    \put(13,-3.5){\small (a)}
    \put(47.5,-3.5){\small (b)}
    \put(83,-3.5){\small (c)}
    }
  \end{overpic}
  \vspace{-3.4mm}
  \caption{
  (a) Input shape. (b) The convex hull $\mathcal{H}$. The vertices are colored blue, and the edges are colored orange. (c) The simplified convex hull $\mathcal{H}'$.
  }
  \label{fig:simplify-convex-hall}
\end{figure}

%
However, a large number of initial control points results in a low optimization speed.
%
Thus, after obtaining the initial control points, we reduce the number of $\{\mathbf{p}_i^h\}$ by simplifying $\mathcal{H}$.
If the number of $\{\mathbf{p}_i^h\}$ in the initialization step is larger than $N_c$, we run the following steps: 
\begin{enumerate}
\item Initialize $\mathcal{H}' = \{\mathbf{v}_i^h\}$.
\item For each vertex $\mathbf{v}_j^h \in \mathcal{H}'$, calculate the Hausdorff distance $d_j$ between the two convex polygons $\mathcal{H}$ and $\mathcal{H}'\setminus \{\mathbf{v}_j^h\}$.
\item If no $d_j$ is smaller than a threshold $\beta$ or the number of vertices in $\mathcal{H}'$ is smaller than $N_c$, stop; otherwise, delete $\mathbf{v}_j^h$ with the minimum $d_j$ from $\mathcal{H}'$, and go to Step (2).
\end{enumerate}
We set $\beta = 2\%d_\text{bb}$ in our experiments.
After reduction, $\mathcal{H}'$ is used to generate the control points $\{\mathbf{p}_i^h\}$ (see Fig.~\ref{fig:simplify-convex-hall}).

\IncMargin{0.5em}
\begin{algorithm}[t]
	\caption{2D Approximation}\label{alg:pseudo-code}
	\newcommand\mycommfont[1]{\footnotesize\ttfamily\textcolor{blue}{#1}}
	\SetCommentSty{mycommfont}
	\SetKwInOut{AlgoInput}{Input}
	\SetKwInOut{AlgoOutput}{Output}

        \SetKwFunction{Initialization}{Initialization}
        \SetKwFunction{ACP}{AddControlPoints}
        \SetKwFunction{UCP}{UpdateCorrespondingPoints}
        \SetKwFunction{DD}{ComputeDescentDirection}
        \SetKwFunction{LS}{PerformLineSearch}

	\AlgoInput{the outer contour $\mg$, the positive weight $\epsilon$, the termination threshold $\theta$, the maximum iteration number $N_{\max}$}
	\AlgoOutput{the B-spline curve $\mathbf{c}$}
	$\mathbf{c}\leftarrow \Initialization(\mg); k \leftarrow 0; \text{flag} \leftarrow 0 $\;
	\While{$E_{error} > \theta$ \text{and} $k < N_{max}$}{
        \tcc{Apply Newton's method to obtain a descent direction}
        $\mathbf{d}\leftarrow \DD(\mg,\mathbf{c})$\;
        $\mathbf{c}\leftarrow \LS(\mg,\mathbf{d})$\;

        \tcc{Update corresponding parameters of sampling points of $\mg$}
        \If{\text{flag} or $(\Delta E_\text{error} < 5 \times 10^{-2} E_\text{error}$ \text{and} $E_\text{error}> N_g d_\text{bb}/20)$}{
        $\UCP(\mg,\mathbf{c})$\; 
        \text{flag} $\leftarrow 1$\;
        }
        \tcc{Add control points at regions with high approximation errors}
        $\mathbf{c}\leftarrow\ACP(\mathbf{c})$\;
        $\epsilon \leftarrow 0.8\epsilon; k \leftarrow k+1$\;
	}

\end{algorithm}
\DecMargin{0.5em}

\paragraph{Optimization}
We use Newton's method to calculate a descent direction for optimizing the positions of control points.
In line search, the step size will become 0 when some part of the B-spline curve $\mathbf{c}$ is close to the target polygon $\mg$ if we update all control points using the same step size.
This prevents $\mathbf{c}$ from moving.
To address this, we assign different step sizes to different control points.
For each control point, we calculate $E_\text{avg}$, the average of $E_\text{error}$ in their influence interval, and update the control points one by one according to $E_\text{avg}$ in descending order.
%
When updating one control point, barrier functions are not used to ensure collision-free constraints, as they inhibit the B-spline curve from being very close to the shape boundary. 
Instead, we first subdivide the control net of $\mathbf{c}$ $k$ times ($k=8$ in our experiments), then perform the continuous collision detection 
algorithm~\cite{redon2002fast} between the subdivided control net and $\mg$ to compute the maximum allowable step size $t_\text{max}$ for the control point to avoid collision.
%
%
Then, we perform a standard Armijo backtracking algorithm with an initial step size $0.5t_\text{max}$ to search the final step size. 

\paragraph{Adding new control points}
The number of control points significantly impacts the B-spline curve's approximation power.
Thus, we adaptively add new control points to the segments with high approximation error to reduce it. 
%
It runs in each iteration as follows: 
(1) sorting the segments of $\mathbf{c}$ based on the average of $E_\text{error}$ in each segment, 
(2) selecting the segments with the top three largest $E_\text{avg}$ from the set of segments with lengths greater than a threshold $l_\text{min}$, and 
(3) if the number of selected segments in Step (2) is zero or the total control points is greater than $N_\text{total}$, stop; otherwise, add $N_\text{add}$ control points uniformly in the selected segments.
We find that gradually adding control points in each iteration can improve efficiency (about threefold) compared to adding them when the algorithm cannot reduce the approximation error.
The former strategy leads to a significant reduction in the number of iterations in practice.
In our experiments, we set $l_\text{min} = 0.25\%d_\text{bb}$, $N_\text{total} = 120$, and $N_\text{add} = 3$.

%
%
%
%
%
%
%
%
%
%

\begin{figure}[t]
  \centering
  \begin{overpic}[width=1.0\linewidth]{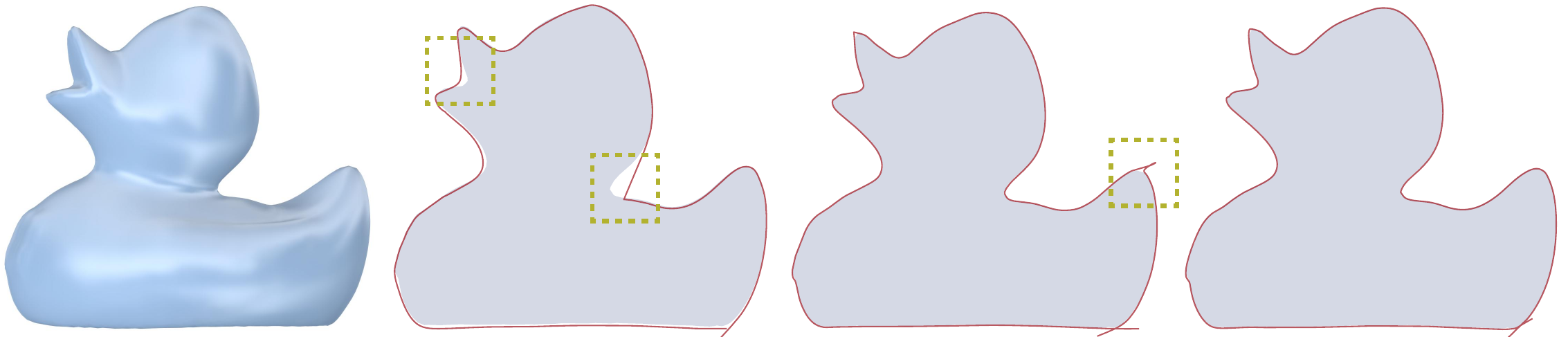}
    {
    \put(10,-3.5){\small \textbf{(a)}}
    \put(35,-3.5){\small \textbf{(b)}}
    \put(60,-3.5){\small \textbf{(c)}}
    \put(85,-3.5){\small \textbf{(d)}}
    }
  \end{overpic}
  \vspace{-3.4mm}
  \caption{
  An ablation study of our corresponding point updating method.
(a) Input shape. (b) Only using the initial correspondence without updating. (c) Only using the closest point as the corresponding point. (d) Our updating.  }
  \label{fig:bad-2D-fitting-result} 
\end{figure}
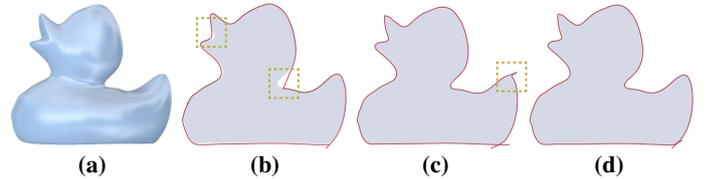

\paragraph{Updating corresponding points}
At the beginning of the optimization, for each sampling point $\mathbf{v}_i$, we use the ratio between the boundary length from $\mathbf{v}_i$ to $\mathbf{v}_0^h$ and the total boundary length to determine its corresponding $u_i$ and set $\mathbf{c}(u_i)$ as its corresponding points.
%
%
After optimization, there are still areas with a high approximation error, specifically for the complex polygon $\mg$ (see Fig.~\ref{fig:bad-2D-fitting-result} (b)).
We then replace our distance function with the local approximation function proposed by~\cite{pottmann2002approximation} and use the closest point in $\mathbf{c}$ as the corresponding point to continue optimizing $\mathbf{c}$.
In practice, when the relative change in the objective function is less than $5 \times 10^{-2}$ and $E_\text{error}>N_g d_\text{bb}/20$ (where $N_g$ indicates the number of sampling points in $\mg$), we start recomputing the corresponding points and then update the corresponding points in each subsequent iteration.
Fig.~\ref{fig:bad-2D-fitting-result} shows an ablation study of updating corresponding points. 

%
%
%
%
%
%

\paragraph{Termination}
%
%
%

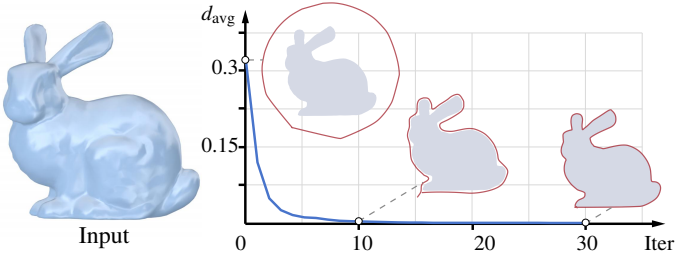
\begin{figure}[t]
  \centering
  \begin{overpic}[width=0.99\linewidth]{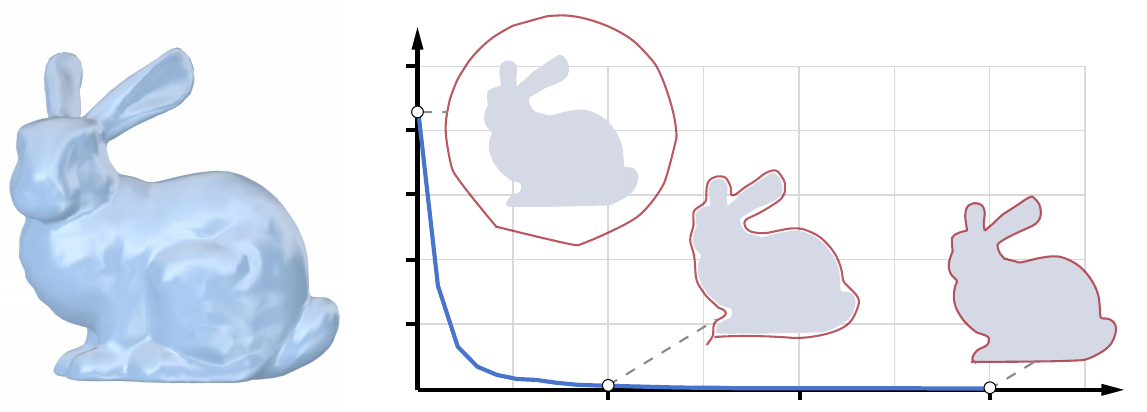}
    {
 \put(35,-2){\footnotesize 0}
 \put(52,-2){\footnotesize 10}
 \put(70,-2){\footnotesize 20}
 \put(85,-2){\footnotesize 30}

 \put(95,-2){\footnotesize Iter}

 \put(30,12.5){\footnotesize 0.15}
  \put(31.5,24){\footnotesize 0.3}

 \put(30,32){\footnotesize $d_\text{avg}$}
 \put(12,-1){\small Input}
}
  \end{overpic}
  \vspace{-1mm}
  \caption{
 Input shape (left). The bidirectional average distance $d_\text{avg}$  between $\mathbf{c}$ and $\mathbf{g}$ change during optimization (right). 
  }
  \label{fig:energy_change}
\end{figure}

The algorithm stops when the approximation error $E_\text{error}$ is less than $10^{-6}N_g$ or when the number of iterations exceeds the specified number $N_\text{max}$ (30 in our experiments).
%
%
Fig.~\ref{fig:energy_change} shows an example of our 2D approximation algorithm.
We use $d_\text{avg}$, the bidirectional average distance, to measure the quality of our approximation result, which is defined as follows:
\begin{equation}\label{equ:d_avg}
  d_\text{avg} = \sum_{\mathbf{v}_i \in S_g} \frac{d(\mathbf{v}_i, \mathbf{c})}{2N_g} +  \sum_{\mathbf{c}(u_i) \in S_c} \frac{d(\mathbf{c}(u_i), \mathbf{g})}{2N_c},
\end{equation}
where $d(x, y)$ is the closest distance between $x$ and $y$, $N_c$ and $N_g$ are the number of $S_c$ and $S_g$ respectively.
 

\subsubsection{Ruled surface extrusion}\label{sec:extrusion}
After the 2D approximation step, we obtain a smooth curve $\mathbf{c}$ close to the outer contour line without colliding it.
We use $\mathbf{c}$ as the base curve of the ruled surface and perform the following steps to construct the surface:
\begin{enumerate}
    \item Set the depth value of each point on $\mathbf{c}$ to be $r$ and convert $\mathbf{c}$ from the camera coordinate system to the world coordinate system.
    \item Translate $\mathbf{c}$ one along the positive and negative camera viewpoint directions to generate two B-spline curves $\mathbf{c}_0$ and $\mathbf{c}_1$. 
    \item Construct a ruled surface $\mathbf{s}(u, v) = (1-v)\mathbf{c}_0(u) + v\mathbf{c}_1(u)$, where $u, v\in [0, 1]$. 
\end{enumerate}
We attempt to optimize the 3D ruled surface $\mathbf{s}$ to further reduce the approximation error; however, $\mathbf{s}$ remains unchanged after optimization. 
Therefore, we abandon this 3D optimization step (see Section~\ref{sec:evaluations}).

\section{Experiments and evaluations}\label{sec:results}
We set the terminating condition of our method as $\alpha=0.025$ and $N_\text{iter}=15$.
We have applied our algorithm to various models to test its performance.
Besides 25 simulated results, we also fabricate 10 physical fabrication examples.
Our method is implemented in C++, and we execute our experiments on a desktop PC with a 3.80 GHz Intel Core i7-10700 and 32GB of memory.

\subsection{Algorithm evaluations}\label{sec:evaluations}


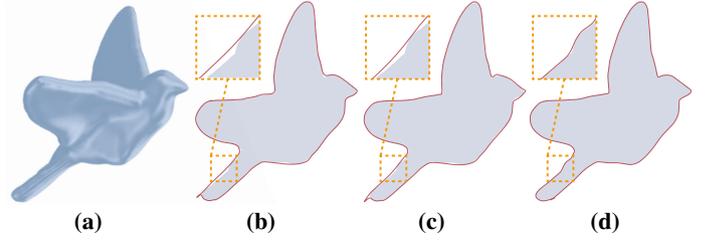
\begin{figure}[t]
  \centering
  \begin{overpic}[width=1.0\linewidth]{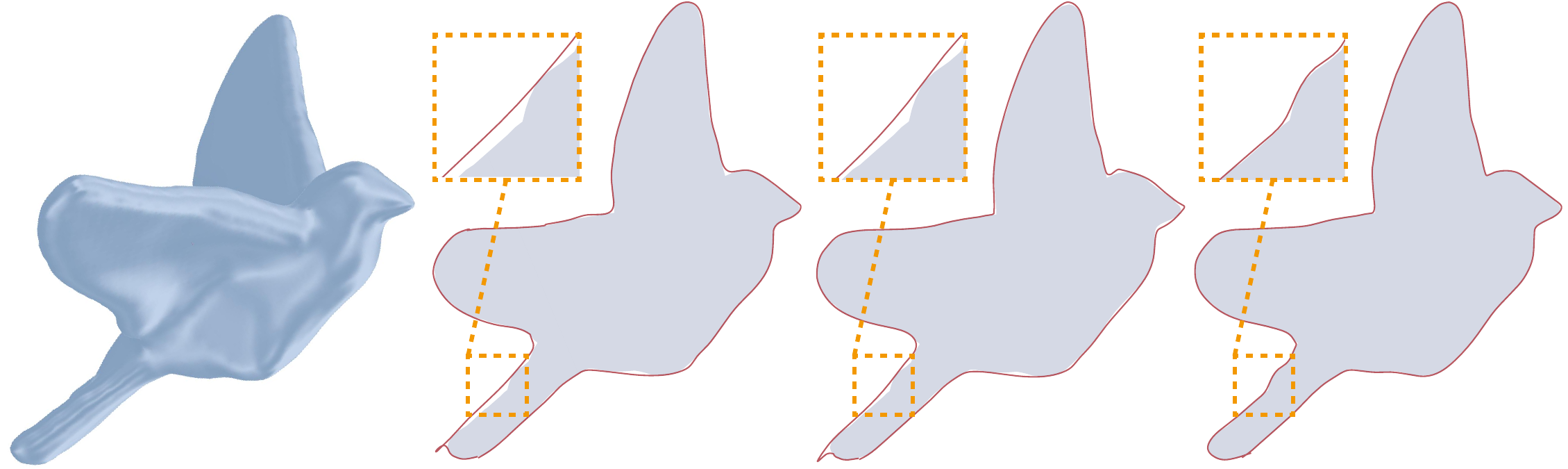}
    {
    \put(10,-3.5){\small \textbf{(a)}}
    \put(35,-3.5){\small \textbf{(b)}}
    \put(60,-3.5){\small \textbf{(c)}}
    \put(85,-3.5){\small \textbf{(d)}}
    }
  \end{overpic}
  \vspace{-3.4mm}
  \caption{
 The ablation study on $\delta_i$ and $\epsilon$.(a) Input shape. (b) Only use $\delta_i$ ($d_\text{avg}=1.51\times10^{-3}$). (c) Only use $\epsilon$ ($d_\text{avg}=1.17
\times10^{-3}$). (d) Our result ($d_\text{avg}=7.46\times10^{-4}$).  
  }
  \label{fig:progressive}
\end{figure}

\paragraph{Ablation study on $\delta_i$ and $\epsilon$}
In $E_\text{error}$, we use $\delta_i$ and $\epsilon$ to improve the approximation error term based on point distance for better approximation results.
Fig.~\ref{fig:progressive} shows the ablation study.
We remove any one of $\delta_i$ and $\epsilon$ from $E_\text{error}$ to perform the optimization (see Fig.~\ref{fig:progressive} (b) and (c)).
Since $\delta_i$ removes mismatched correspondences between $\mathbf{c}$ and $\mathbf{g}$, while the diminishing constant $\epsilon$ controls the convergence speed to prevent parts of 
$\mathbf{c}$ from approaching $\mathbf{g}$ too early and blocking the motion of neighboring regions, removing either of the two parameters increases the error of the results (see Fig.~\ref{fig:progressive} (b) and (c)).
In contrast, our algorithm achieved the result with the smallest error (see Fig.~\ref{fig:progressive} (d)).

%
%

\begin{figure}[t]
  \centering
  \begin{overpic}[width=1.0\linewidth]{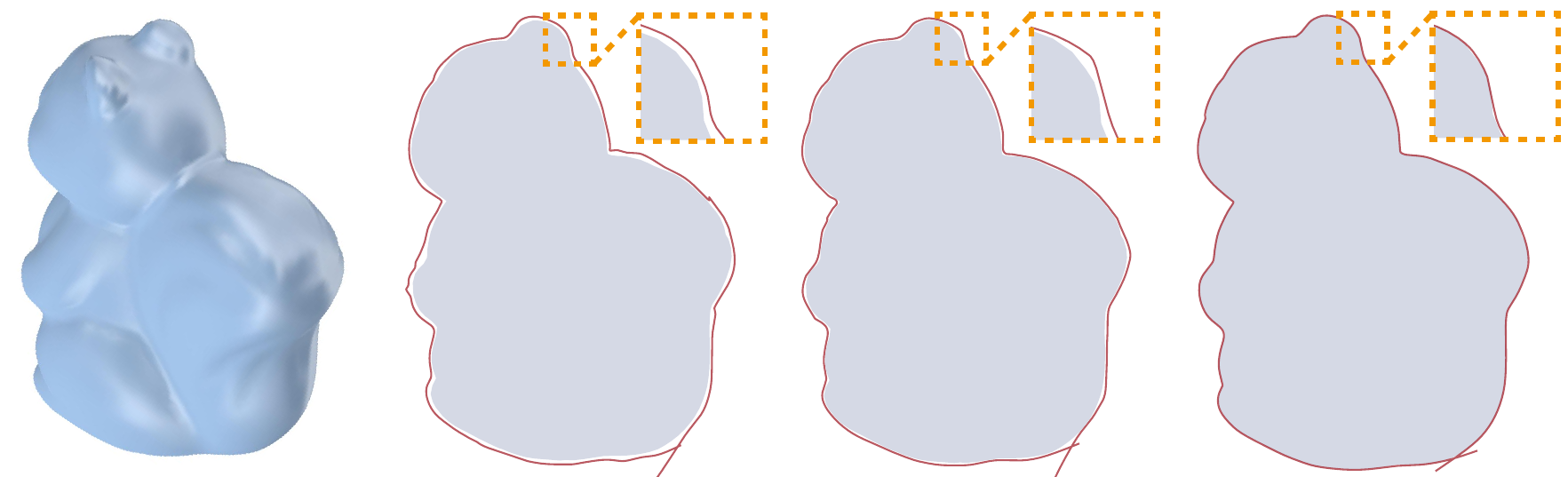}
    {
    \put(10,-3.5){\small \textbf{(a)}}
    \put(35,-3.5){\small \textbf{(b)}}
    \put(60,-3.5){\small \textbf{(c)}}
    \put(85,-3.5){\small \textbf{(d)}}
    }
  \end{overpic}
  \vspace{-3.4mm}
  \caption{
  Comparison with using a barrier function. (a) Input shape. (b) $w_\text{barrier} = 1$, $d_\text{avg}=8.56\times10^{-3}$. (c) $w_\text{barrier} = 0.05$, $d_\text{avg}=6.79\times10^{-3}$. (d) Ours, $d_\text{avg}=1.27\times10^{-3}$. 
  }
  \label{fig:barrier}
\end{figure}

\begin{figure}[t]
  \centering
  \begin{overpic}[width=0.99\linewidth]{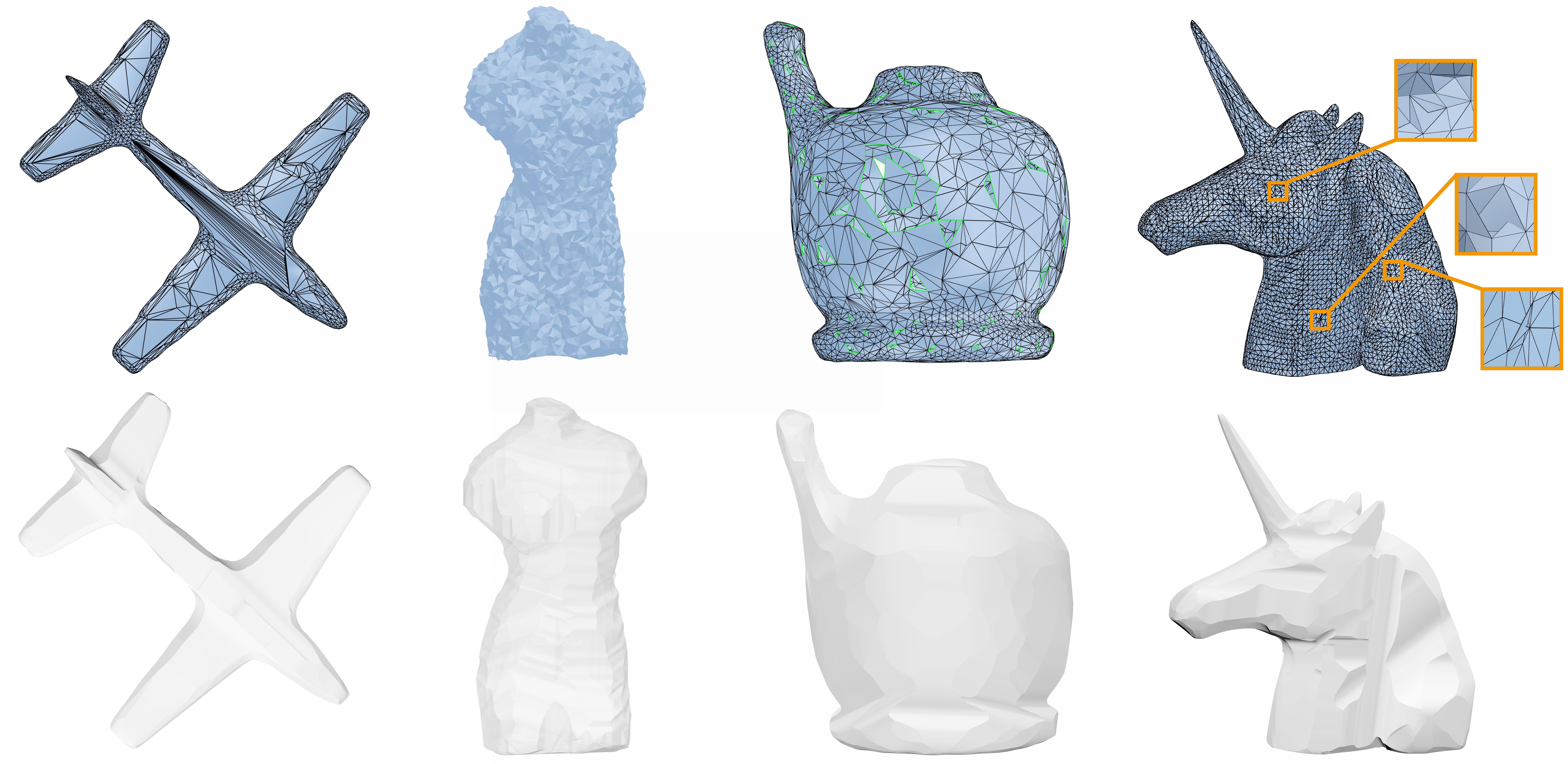}
    {
    \put(5,-3.5){\small Sparse}
    \put(30,-3.5){\small Noisy}
    \put(55,-3.5){\small Non-watertight}
    \put(80,-3.5){\small
 Non-manifold}
    }
  \end{overpic}
  \vspace{0.5mm}
  \caption{
        Stress test. 
        From left to right, the number of cuts for each model is 8, 8, 11, and 12, respectively.
        Correspondingly, the average errors are $6.28\times10^{-3}$, $1.02\times10^{-2}$, $6.42\times10^{-3}$ and $8.77\times10^{-3}$, respectively.
  }
  \label{fig:more-test}
\end{figure}

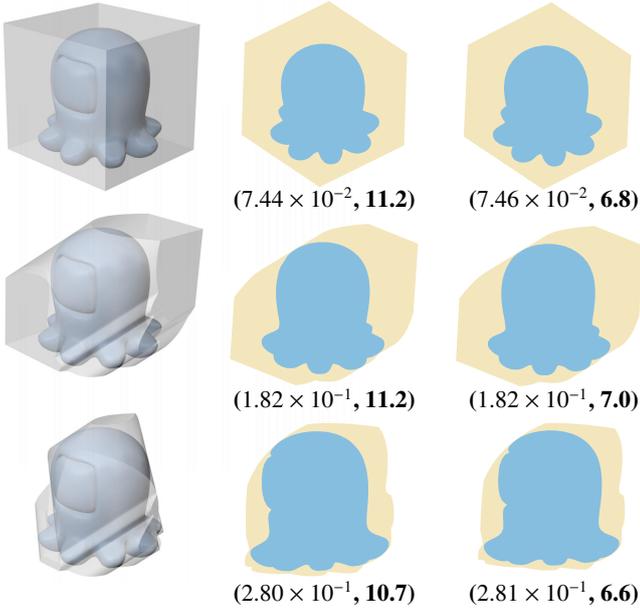
\begin{figure}[t]
  \centering
  \begin{overpic}[width=0.95\linewidth]{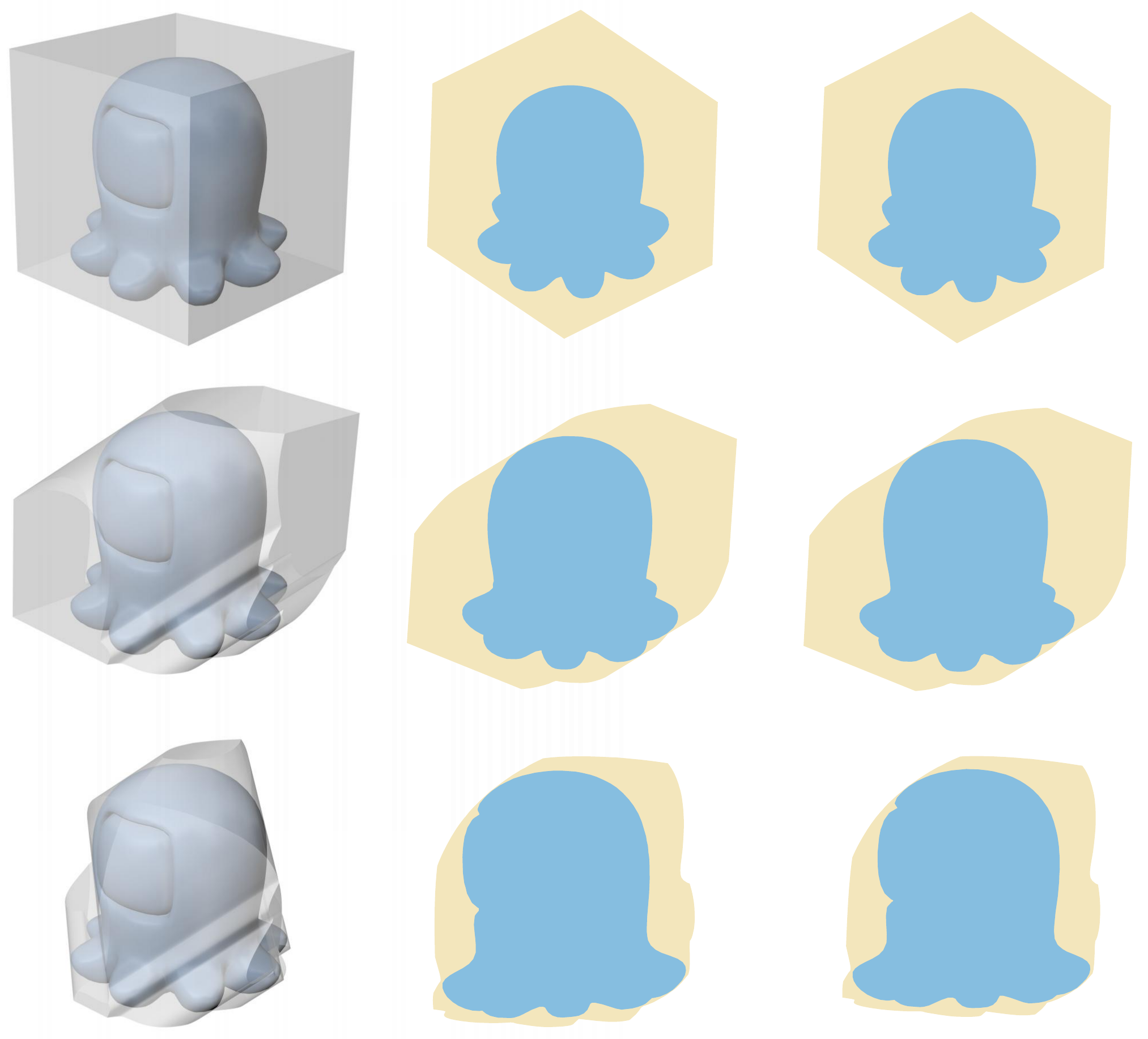}
    {
    \put(36,-2.7){\small \textbf{($2.80\times10^{-1}$, 10.7)}}
    \put(72,-2.7){\small \textbf{($2.81\times10^{-1}$, 6.6)}}
    
    \put(36,27){\small \textbf{($1.82\times10^{-1}$, 11.2)}}
    \put(72,27){\small \textbf{($1.82\times10^{-1}$, 7.0)}}
    
    \put(36,58){\small \textbf{($7.44\times10^{-2}$, 11.2)}}
    \put(72,58){\small \textbf{($7.46\times10^{-2}$, 6.8)}}


    }
  \end{overpic}
  \caption{
    Genetic algorithm vs the sampling method. We evaluate our method on the octopus model by applying materials of different shapes. 
    The first column shows the input, the second column presents the results obtained using the sampling-based method, and the third column displays the results generated by our algorithm.
    The text below each result indicates $E^\text{ours}_\text{material}$ and time consumption (seconds).}
    \label{fig:cmp_sampling}
\end{figure}

\paragraph{Collision-free constraint}
As we keep the collision-free constraint during optimization, adding a barrier function with the weight $w_\text{barrier}$ is another straightforward way to achieve this goal.
However, our experiments find that the barrier function prevents the B-spline curve from being very close to the shape boundary. 
Fig.~\ref{fig:barrier} shows a comparison between using a $-\log$ barrier function and our method.
The smaller the weight of the barrier function is, the smaller the approximation error is  (see Fig.~\ref{fig:barrier} (b) and (c)).
%
However, it is still greater than that of ours (see Fig.~\ref{fig:barrier} (d)).
In addition, when the distance is very small, the barrier function will numerically explode, increasing the error.
In contrast, our results are close enough to the boundary.


\paragraph{Stress test}
We test our algorithm on the non-uniform triangulation, noisy, non-water-tight, and non-manifold meshes to prove its practicality (Fig.~\ref{fig:more-test}).
%
%
We obtain small approximation errors and a few cuts for these models, which are difficult to handle by the mesh-based methods.
These results prove that our algorithm is not affected by the mesh quality and tessellation and is practical and effective. 

\paragraph{Genetic algorithm vs the sampling method}
We compare our genetic algorithm with a sampling-based view selection approach.
For the sake of fairness, we only replaced the viewpoint selection part in our algorithm while keeping the rest unchanged.
The sampling-based method is performed on a uniformly sampled viewpoint set with 1000 candidate viewpoints.
The results are shown in Fig.~\ref{fig:cmp_sampling}.
For each model, our method yields a comparable area error to the sampling-based approach, but reduces computation time by nearly $60\%$, demonstrating significantly improved efficiency.
%
%
%

\begin{figure}[t]
  \centering
  \begin{overpic}[width=1.0\linewidth]{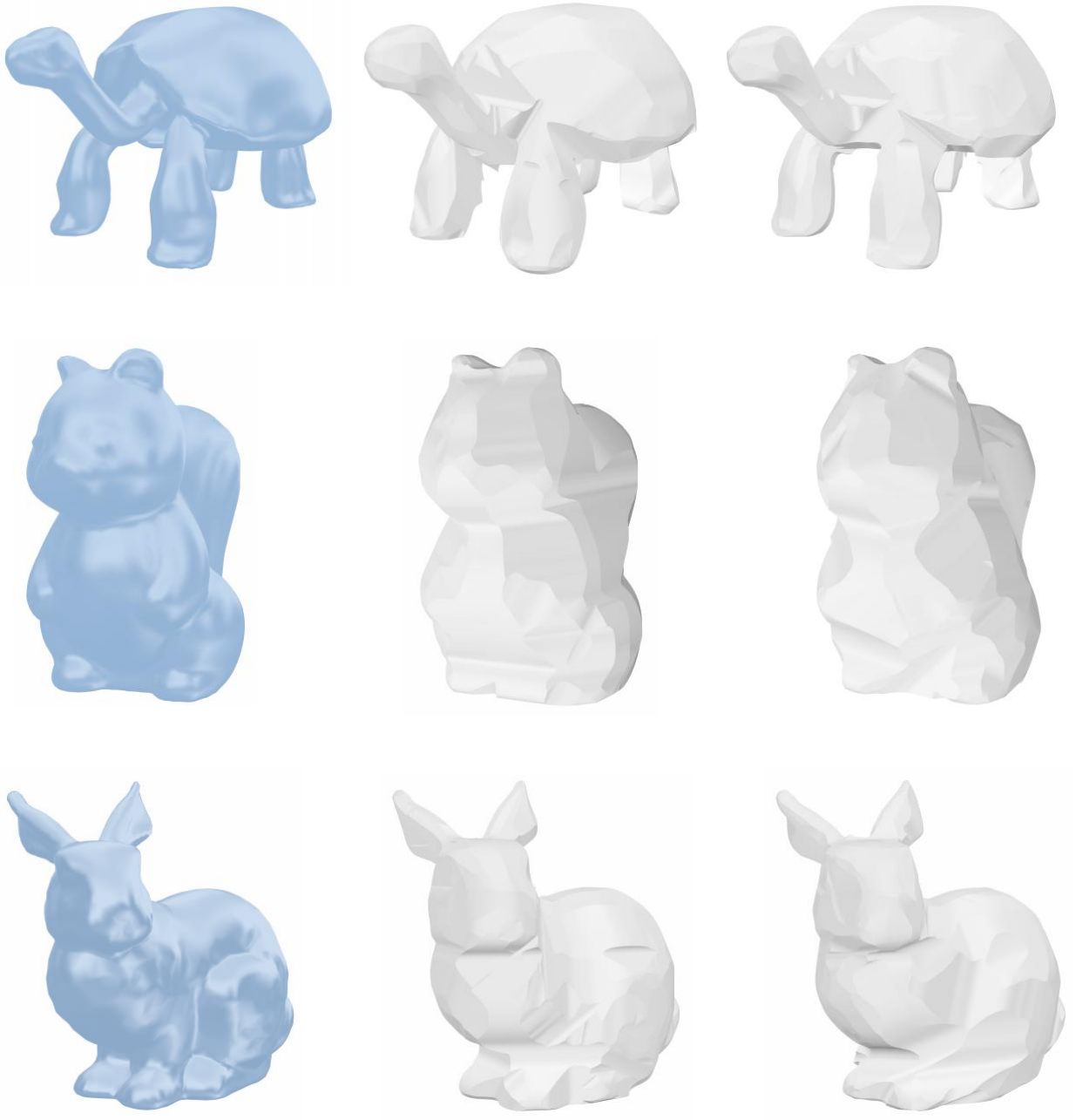}
    {
     \put(10,71){\small Tortoise}
    \put(36,71){\small \textbf{($1.17\times10^{-2}$, 8, 82.1)}}
    \put(71,71){\small \textbf{($6.96\times10^{-3}$, 8, 28.8)}}

    \put(7,33){\small Squirrel}
    \put(35,33){\small \textbf{($1.33\times10^{-2}$, 10, 87.3)}}
    \put(70,33){\small \textbf{($9.28\times10^{-3}$, 10, 33.1)}}
    
    \put(10,-3){\small Bunny}
    \put(36,-3){\small \textbf{($1.13\times10^{-2}$, 15, 135.6)}}
    \put(71,-3){\small \textbf{($7.97\times10^{-3}$, 15, 56.8)}}
    }
  \end{overpic}
  \vspace{-3.4mm}
  \caption{
 Comparisons with practitioners.  
 The second and third columns show the practitioner's and our results, respectively.
The text below each result indicates the average error, number of cuts, and time consumption (minutes).
  }
  \label{fig:Comparisons}
\end{figure}

\paragraph{Comparisons with practitioners}
We compare with a practitioner using Rhinoceros 3D software~\cite{rhinoceros3d} to generate the cutting paths on three models (Fig.~\ref{fig:Comparisons}).
%
For each model, we limit the number of the practitioner's cuts to the same as ours to compare the average distance from results to the input shapes and the total time spent.
%
Our approximation error and running time are both smaller than those of practitioners for each model.

\paragraph{Comparisons with the differentiable rendering}

\begin{figure}[t]
  \centering
  \begin{overpic}[width=0.99\linewidth]{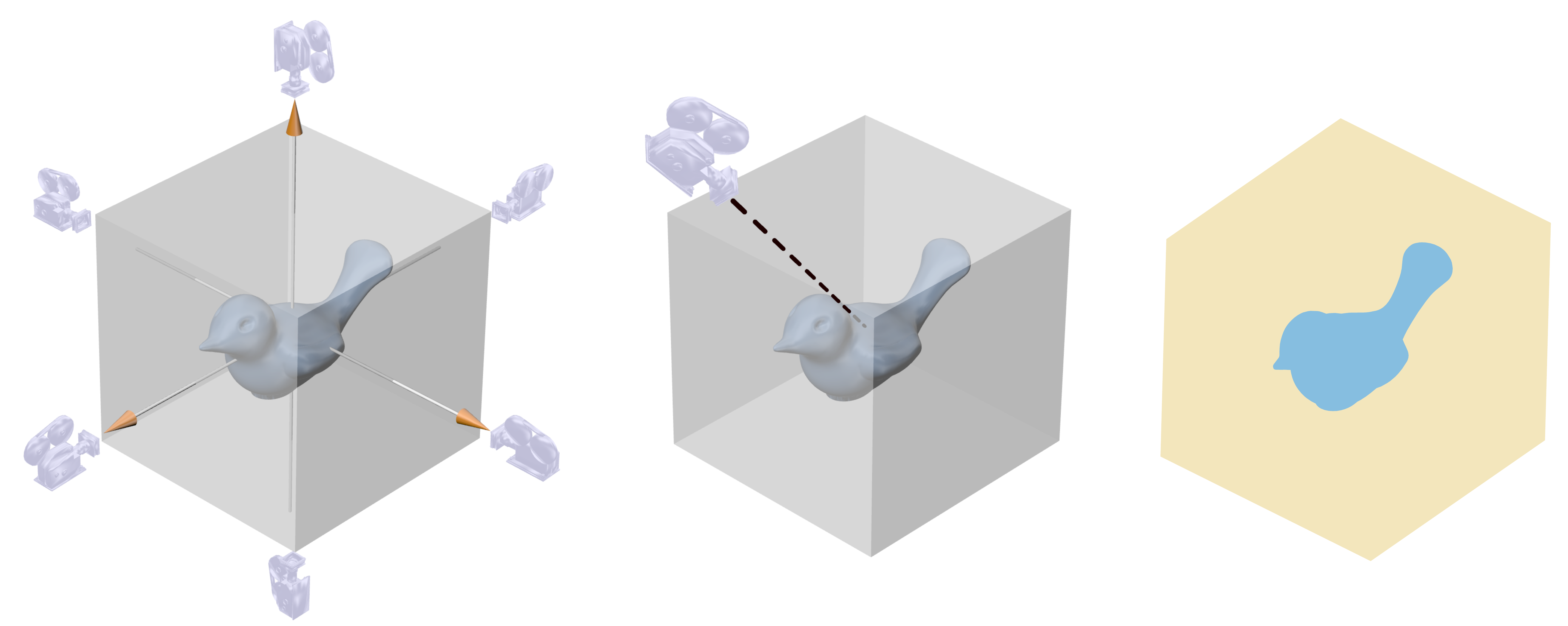}
    {
 \put(18,-3.5){\small \textbf{(a)}}
 \put(54,-3.5){\small \textbf{(b)}}
 \put(85,-3.5){\small \textbf{(c)}}
    }
  \end{overpic}
  \vspace{1mm}
  \caption{
  Our viewpoint optimization method begins with six initial viewpoints (a) to find the viewpoint with the minimum energy (b). The rendering image of the optimal viewpoint is shown (c).
  }
  \label{fig:view-initialization}
\end{figure}

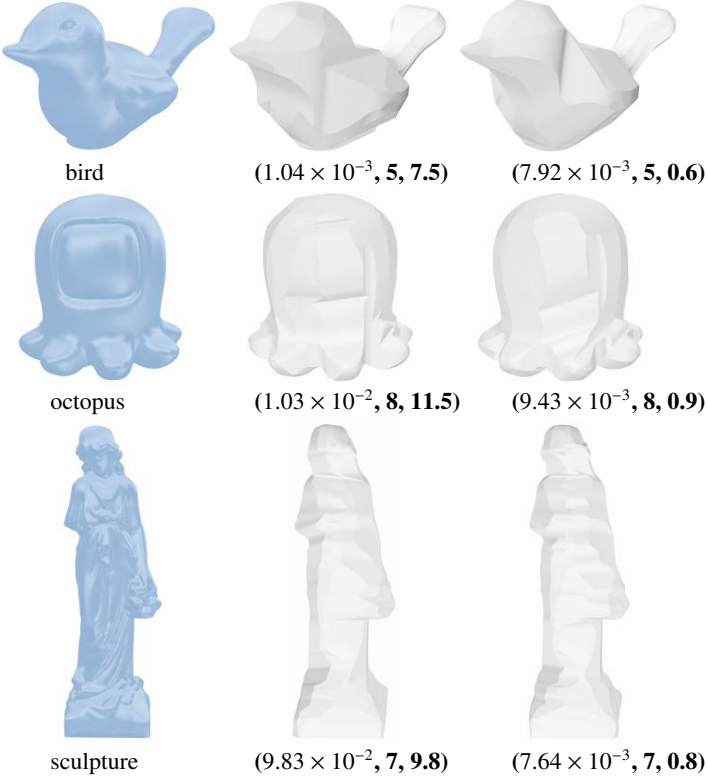
\begin{figure}[t]
  \centering
  \begin{overpic}[width=0.99\linewidth]{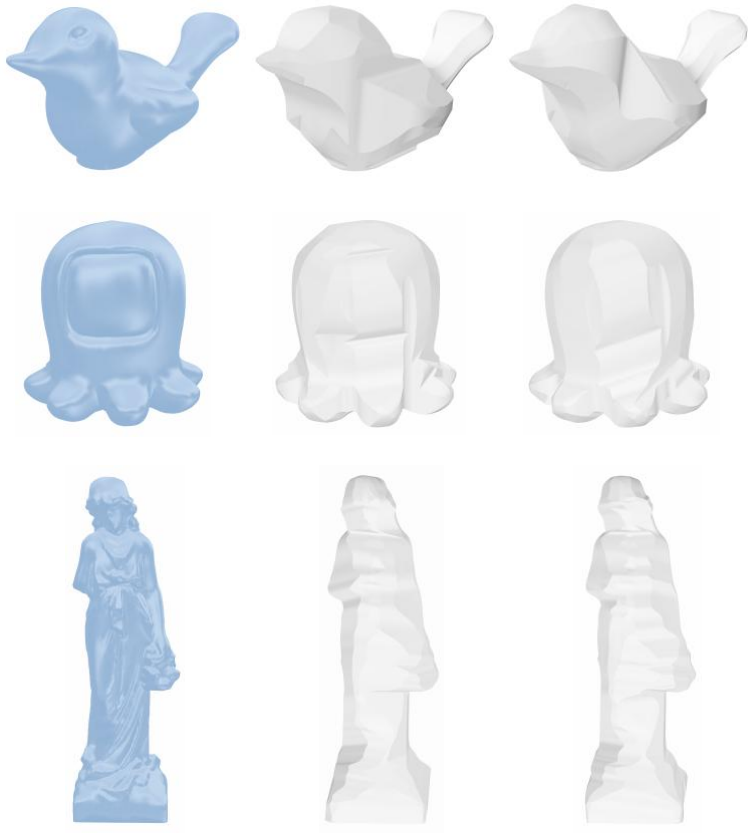}
    {
    \put(7,-3){\small sculpture}
    \put(34,-3){\small \textbf{($9.83\times10^{-2}$, 7, 9.8)}}
    \put(68,-3){\small \textbf{($7.64\times10^{-3}$, 7, 0.8)}}

    \put(7,45){\small octopus}
    \put(34,45){\small \textbf{($1.03\times10^{-2}$, 8, 11.5)}}
    \put(68,45){\small \textbf{($9.43\times10^{-3}$, 8, 0.9)}}

    \put(9,75.5){\small bird}
    \put(34,75.5){\small \textbf{($1.04\times10^{-3}$, 5, 7.5)}}
    \put(68,75.5){\small \textbf{($7.92\times10^{-3}$, 5, 0.6)}}
    
    }
  \end{overpic}
  \vspace{1mm}
  \caption{
Comparisons with the differentiable rendering. 
The second and third columns show the differentiable rendering results and our results, respectively. The text below each
result indicates the average error, number of cuts, and viewpoint optimization time consumption (minutes).
%
  }
  \label{fig:Baseline}
\end{figure}

To validate the effectiveness of our view selection approach based on a genetic algorithm, we compare a differentiable rendering-based viewpoint optimization method for minimizing $E^\text{ours}_\text{material}$~\eqref{equ:our-viewpoint-obj}.

We use \textit{SoftSilhouetteShader} based on SoftRas~\cite{liu2019soft} to make the objective function $E^\text{ours}_\text{material}$ differentiable.
The $\sigma$ and $\gamma$ of SoftRas in the shader are set to $10^{-5}$ and $10^{-5}$, respectively.
Other parameters in the shader are the default.
Samely, we use the orthogonal projection to render $256\times256$ images and limit the camera's movement range to a sphere surface with a radius of 2 and a center located at the origin.
%
%
In this case, we represent the camera position $(x,y,z)$ as follows:
\begin{equation}\label{equ:camera-pos}
\left\{
\begin{aligned}
x & =  r \cos(k \phi + \phi_\text{init}) \cos(k \theta + \theta_\text{init}) \\
y & =  r \cos(k \phi + \phi_\text{init}) \sin(k \theta + \theta_\text{init}) \\
z & =   r \sin(k \phi + \phi_\text{init})
\end{aligned}
\right.
,
\end{equation}
where $k$ is positive to control the severity of angle changes, and $\phi_\text{init}$ and $\theta_\text{init}$ are their initial values.
In our experiments, we set $k=5$.

Since the objective function $E^\text{ours}_\text{material}$ is highly non-linear, we use 6 different initial values to solve the problem for a better result.
The initial camera positions are located on the positive and negative half axes of the three coordinate axes.
Fig.~\ref{fig:view-initialization} shows the initialization strategy and one result.
We apply Adam~\cite{Kingma2014AdamAM} with a learning rate 0.1 to solve the viewpoint optimization problem. 
However, this method frequently falls into local optima rather than achieving the global optimum in our experiments.

As shown in Fig.~\ref{fig:Baseline}, we limit the number of cuts in the results obtained by differentiable rendering to the same number as ours to compare the average distance from the results to the input shapes and the time to obtain the optimal viewpoint.
Our approximation error and viewpoint selection's running time are both smaller. 

\begin{figure}[t]
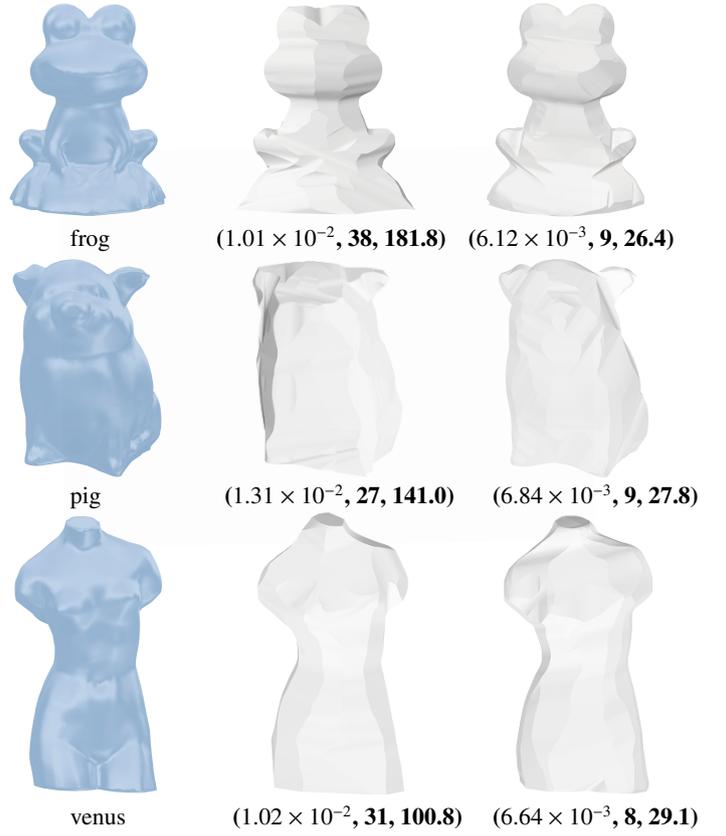

  \centering
  \begin{overpic}[width=0.95\linewidth]{cmp_zz}
    {
    
    \put(8,-2.8){\small venus}
    \put(28,-2.8){\small \textbf{($1.02\times10^{-2}$, 31, 100.8)}}
    \put(60,-2.8){\small \textbf{($6.64\times10^{-3}$, 8, 29.1)}}

    \put(8,37){\small pig}
    \put(27,37){\small \textbf{($1.31\times10^{-2}$, 27, 141.0)}}
    \put(60,37){\small \textbf{($6.84\times10^{-3}$, 9, 27.8)}}

    \put(8,69){\small frog}
    \put(26,69){\small \textbf{($1.01\times10^{-2}$, 38, 181.8)}}
    \put(57,69){\small \textbf{($6.12\times10^{-3}$, 9, 26.4)}}
    
    }
  \end{overpic}
  \vspace{1mm}
  \caption{
Comparisons with Zhang et al.\cite{zhang2025carving}. 
The second and third columns show Zhang et al.\cite{zhang2025carving} and our results, respectively. The text below each result indicates the average error, number of cuts, and time consumption (minutes).
  }
  \label{fig:cmp_zz}
\end{figure}

\paragraph{Comparisons with other methods}
Zhang et al.\cite{zhang2025carving} propose a cover-to-fit strategy. It first generates a small set of cover patches to cover the input shape, considering the property of the ruled surface. Then, perform ruled surface fitting for each patch.
As shown in Fig.~\ref{fig:cmp_zz}, we compared our method with Zhang et al.\cite{zhang2025carving}. For all test cases, our approximation error decreases by about $70\%$, and we use roughly $270\%$ fewer cuts; meanwhile, our runtime is reduced by approximately $414\%$. These results prove the practicality and effectiveness of our method.

\paragraph{Optimizing ruled surfaces}
Given the ruled surface $\mathbf{s} = (1-v) \mathbf{c}_{0}(u) + v \mathbf{c}_{1}(u)$ extruding from $\mathbf{c}$, we aim to use $\mathbf{s}$ to approximate the input shape $\mathcal{M}$ with the collision-free constraint.
Thus, this 3D approximation problem can be formulated as follows:
\begin{equation}\label{equ:3D-objective}
   \min E_\text{error}^\text{3D} + E_\text{smooth}^\text{3D},
\end{equation}
where $E_\text{error}^\text{3D}$ measures the approximation error and $E_\text{smooth}^\text{3D}$ is the smoothness term.

We use the sum of the square of the distance from the sampling points on $s$ to their closest points on $\mathcal{M}$ as our 3D approximation error term:
\begin{equation}\label{equ:3D-approximation-error}
   E_\text{error}^\text{3D} =  \sum_{i=0}^{N} \sum_{j=0}^{M} \|v_{ij} - \mathbf{s}(u_i, v_j)\|_2^2,
\end{equation}
where $\{\mathbf{s}(u_i, v_j)\}$ are the sampling points on $\mathbf{s}$ and $\{v_{ij}\}$ are the closest points on $\mathcal{M}$.
For a ruled surface represented by NUBS, it has u and v two-direction smoothness terms.

Thus, our $E_\text{smooth}^\text{3D}$ is defined as follows:
\begin{equation}
    E_\text{smooth}^\text{3D} =  w_u E_\text{smooth, u}^\text{3D} +  w_v E_\text{smooth, v}^\text{3D},
\end{equation}
where
\begin{equation}\label{equ:3D-smoothness-error}
\begin{split}
    &E_\text{smooth, u}^\text{3D} = \sum_{i=0}^{N} \sum_{j=0}^{M} \|\frac{\partial^2\mathbf{s}}{\partial u^2}(u_i, v_j)\|_2^2,\\
    &E_\text{smooth, v}^\text{3D} = \sum_{i=0}^{N_c - 1} \|(\mathbf{p}_{i,0} - \mathbf{p}_{i,1}) - (\mathbf{p}_{i+1,0} - \mathbf{p}_{i+1,1})\|_2^2.
\end{split}
\end{equation}
$w_u$ and $w_v$ are positive weights.
$N_c$ is the number of the control points of $\mathbf{c}$, and $\{\mathbf{p}_{i, 0}\}$ and $\{\mathbf{p}_{i, 1}\}$ are the control points of $\mathbf{c}_0$ and $\mathbf{c}_1$ respectively.

Similar to our 2D approximation method, we assign different step sizes to different control points, and the updating order is determined by $E_\text{error}^\text{3D}$ in their influence interval.
We subdivide $\mathbf{s}$ in $u$ and $v$ directions $k=8$ and $l=7$ times and then perform the CCD algorithm to calculate the maximum allowable step size as the initial step size in the line search step.
Newton's method is applied to solve our 3D approximation problem.
We set $w_u=1\times 10^{-8}$, $w_v=1\times 10^{-4}$, $N=500$, and $M=50$ in our experiments.
Our method terminates when the relative change of the objective energy is less than $1\times10^{-4}$, or the number of iterations reaches 20.

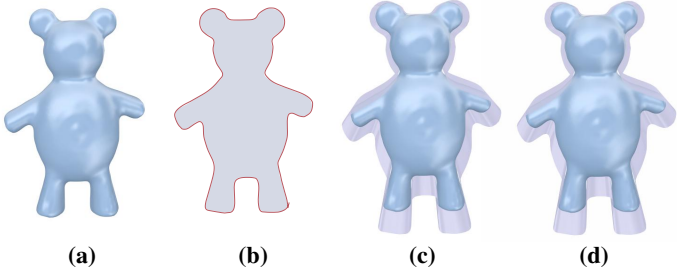
\begin{figure}[t]
  \centering
  \begin{overpic}[width=0.99\linewidth]{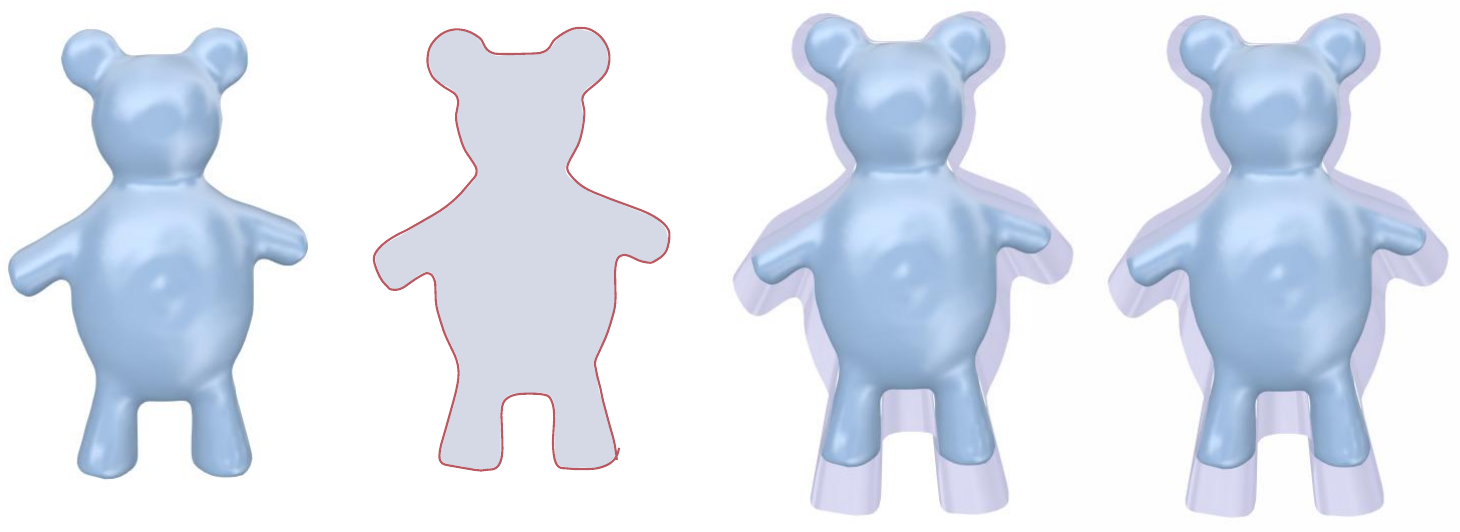}
    {
    \put(10,-3.2){\small \textbf{(a)}}
    \put(35,-3.2){\small \textbf{(b)}}
    \put(60,-3.2){\small \textbf{(c)}}
    \put(85,-3.2){\small \textbf{(d)}}
    }
  \end{overpic}
  \vspace{1mm}
  \caption{
The result of 3D approximation. (a) Input shape. (b) The B-spline approaching the outer contour line (red). (c) The ruled surface extruded from the B-spline (light purple), $d_\text{avg}^\text{3D}=9.25\times10^{-2}$. (d) The optimized ruled surface,  $d_\text{avg}^\text{3D}=9.25\times10^{-2}$.
  }
  \label{fig:3D-results}
\end{figure}

We perform a test to optimize the ruled surface $\mathbf{s}$ to further reduce the approximation error.
Fig.~\ref{fig:3D-results} shows the result.
Since the initial $\mathbf{s}$ is close enough to $\mathcal{M}$, moving any control point of $\mathbf{s}$ for the approximation error reduction 
has a very high probability to trigger collision between $\mathbf{s}$ and $\mathcal{M}$.
%
So, $d_\text{avg}^\text{3D}$, the average of the average distance from the ruled surface to the shape and from the shape to the ruled surface, remains unchanged after the optimization.
Besides, the fabrication error is lower when cutting along cylindrical surfaces, and if $\mathbf{s}$ is a general ruled surface, self-locking may occur between the cut material and the remaining material.
Therefore, we abandon the 3D approximation optimization step from our algorithm.

\paragraph{Timings}
For the Bunny model in Fig.~\ref{fig:rough-machining}, the viewpoint selection and the 2D approximation in each iteration take about 0.1 and 3.5 minutes.
The time of ruled surface extrusion and material removal can be ignored.
In the 2D approximation process, we spend lots of time finding proper step sizes for different control points so that the B-splines satisfy the collision-free constraint and are close enough to the outer contour lines.



\begin{figure*}[!t]
  \centering
  \begin{overpic}[width=0.99\linewidth]{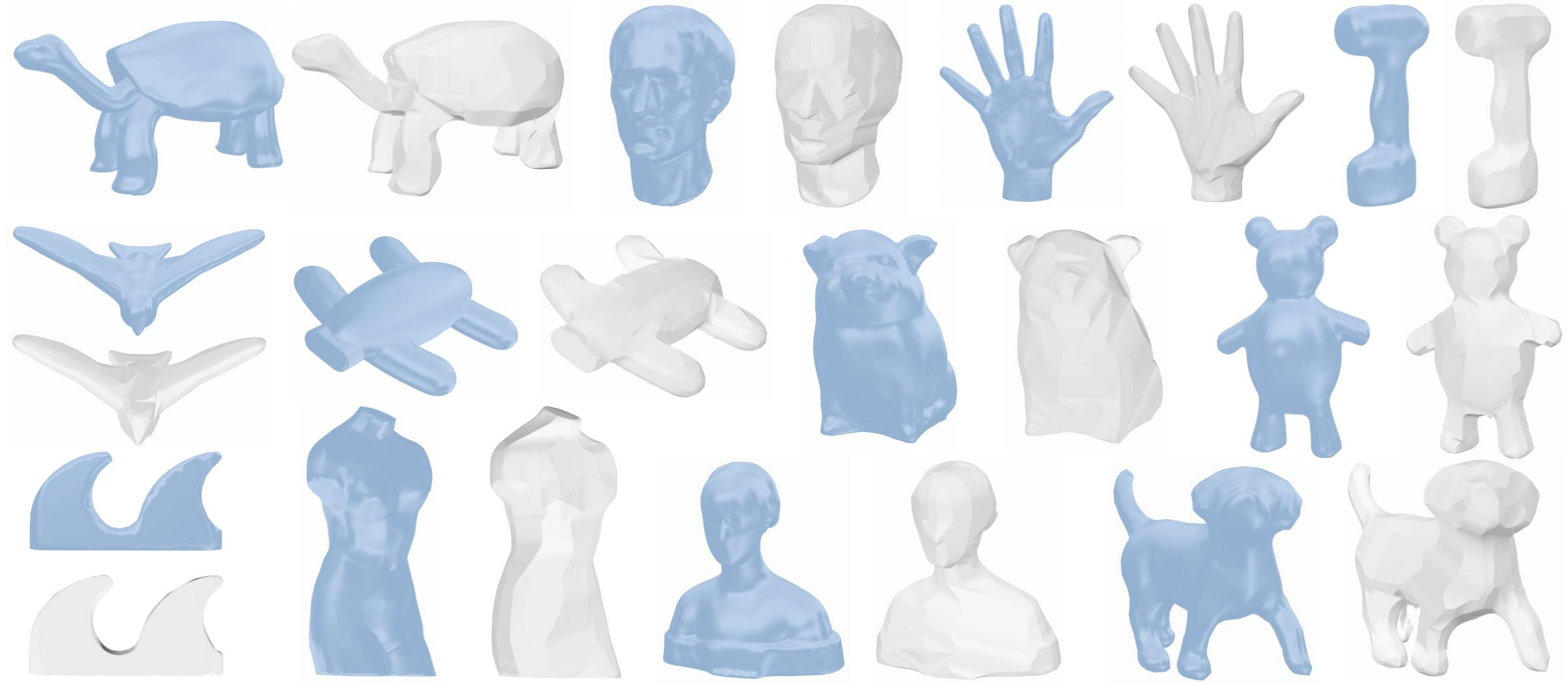}
    {
    }
  \end{overpic}
  \vspace{-2mm}
  \caption{
Twelve examples. For each example, the light blue model is the input, and the white is the result.
  }
  \label{fig:gallery}
\end{figure*}

\begin{figure*}[!t]
  \centering
		\begin{overpic}[width=0.99\linewidth]{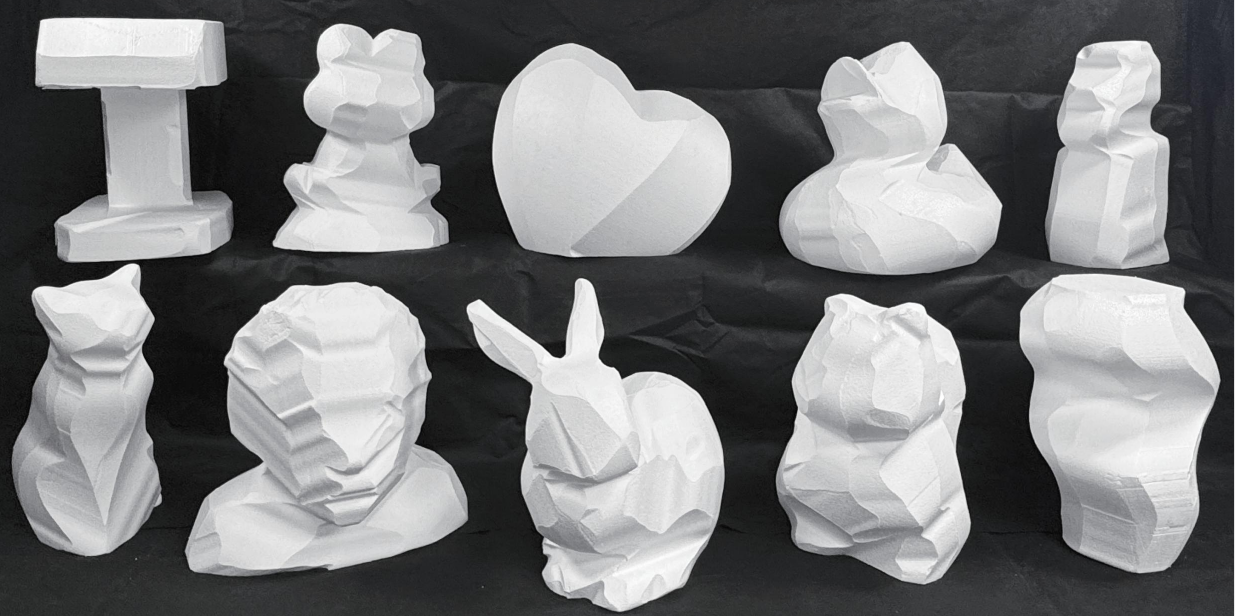}
			{
		    }
		\end{overpic}
		 \vspace{-2mm}
		 \caption{
			All fabrication results.
			 }
		 \label{fig:fab-all}
\end{figure*}

\begin{figure}[t]
  \centering
  \begin{overpic}[width=0.99\linewidth]{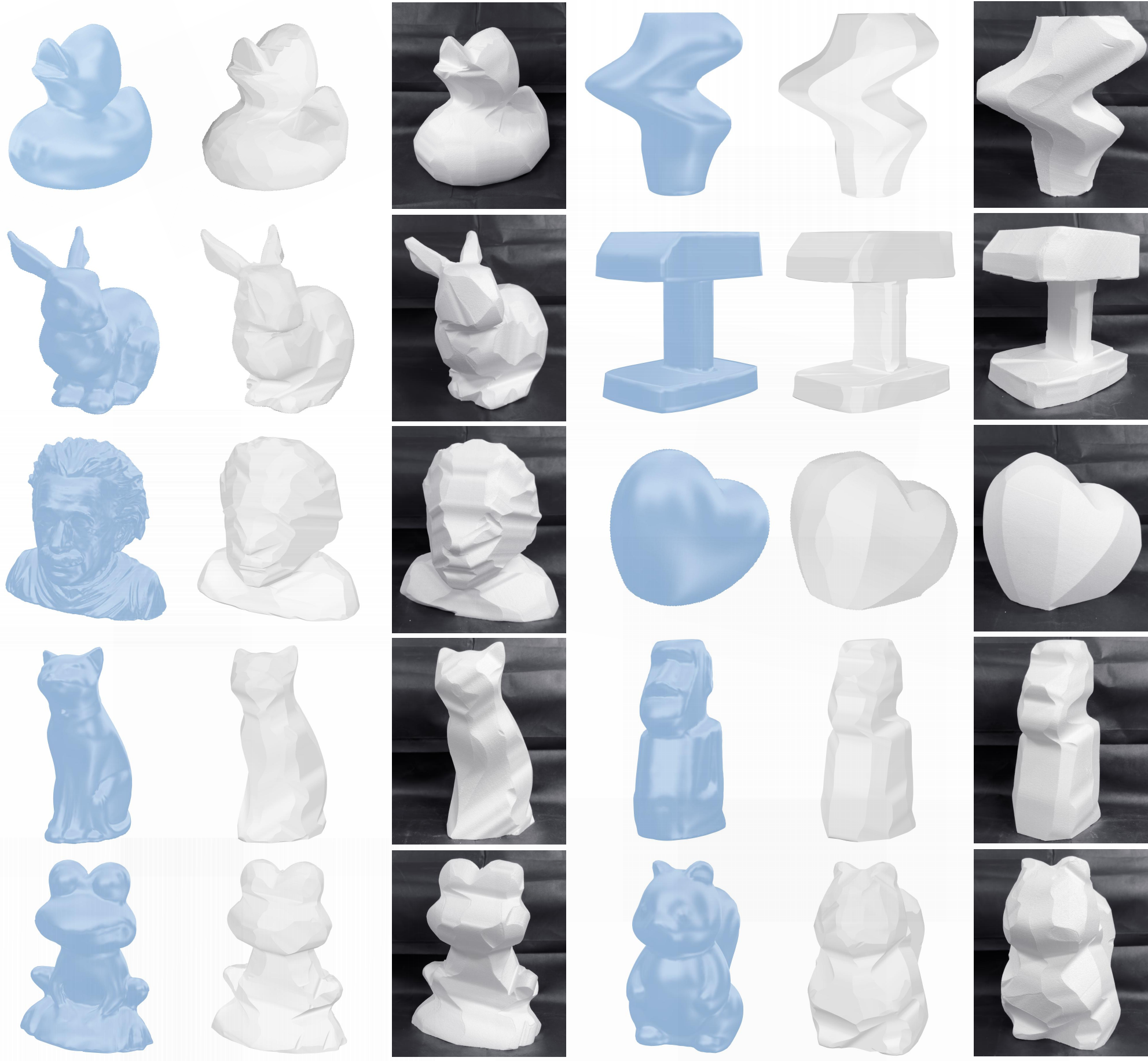}
    {
    }
  \end{overpic}
  \caption{
 Physical fabrication of ten models. For each example, we show the input shape (left), the simulation result (middle), and the fabrication result (right).
  }
  \label{fig:Physical-fabrication}
\end{figure}

\begin{figure}[!htb]
  \centering
  \begin{overpic}[width=0.99\linewidth]{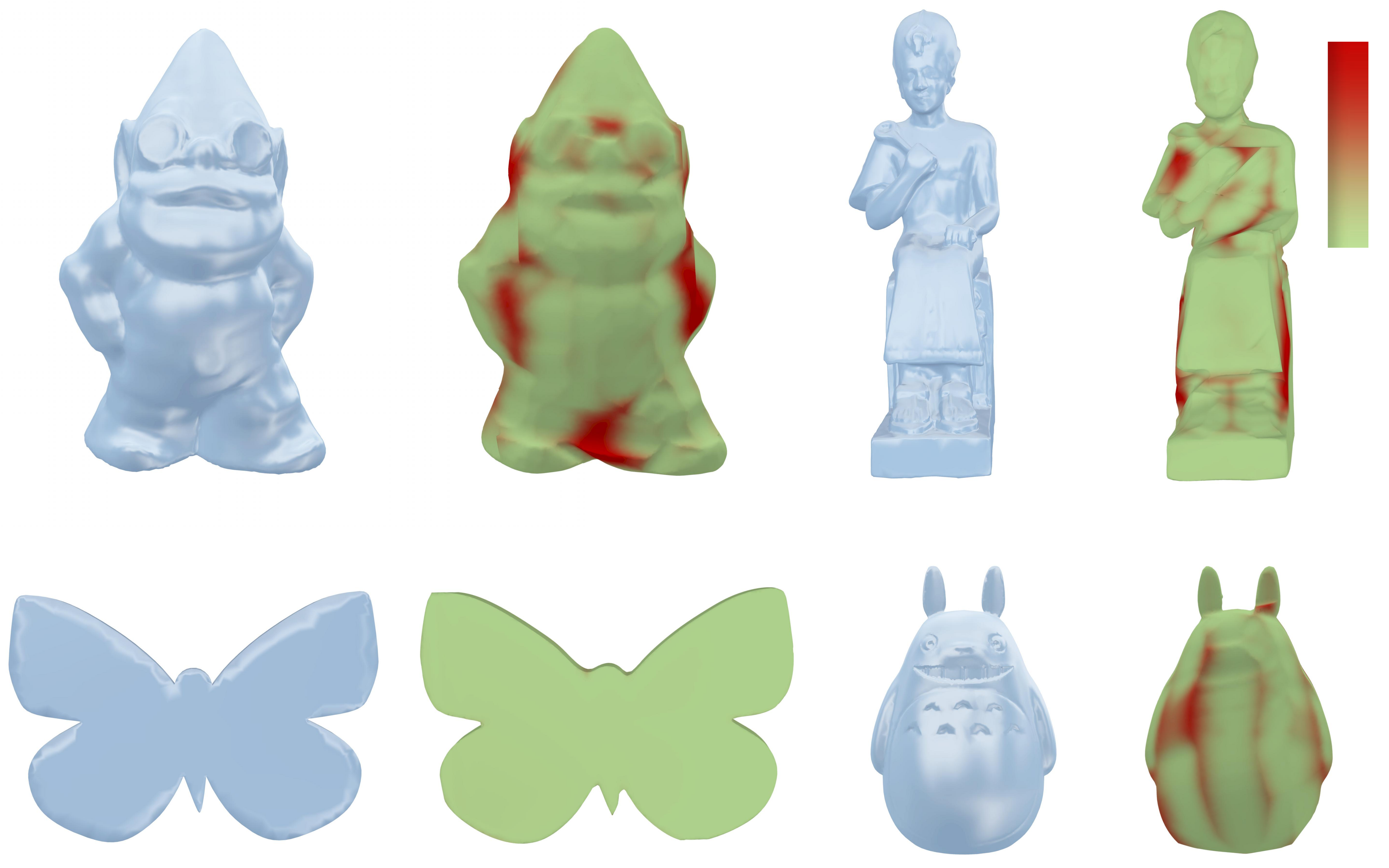}
    {
    \put(19,23.5){\small \textbf{($8.63\times10^{-3}$, 9)}}
    \put(67,23.5){\small \textbf{($7.28\times10^{-3}$, 12)}}
    \put(19,-3.5){\small \textbf{($4.02\times10^{-3}$, 6)}}
    \put(67,-3.5){\small \textbf{($9.02\times10^{-3}$, 5)}}
    \put(96,42){\small \textbf{0}}
    \put(94,60){\small \textbf{0.05}}
    }
  \end{overpic}
  \vspace{1mm}
  \caption{
Simulation error distribution. We use the one-sided distance from the simulation results to the input model to measure the error. The distance map is color-coded by the color bar. The text below each result indicates the average error and number of cuts.
  }
  \label{fig:colormap}
\end{figure}

\paragraph{More examples}
12 simulation results are shown in Fig.~\ref{fig:gallery}.
For these various input shapes, our method successfully generates a small set of ruled surfaces for the rough machining process.
We also use a hot-wire cutting machine to fabricate 10 models (see Figs.~\ref{fig:fab-all},~\ref{fig:Physical-fabrication}).
These simulation and fabrication results prove the practicality and effectiveness of our method.
Although the speed of our method is still not fast enough, it is automatic and does not require manual labor while producing results better than those obtained manually.  
Therefore, our approach holds considerable practical value in digital fabrication.
The other four examples' error distribution maps are shown in Fig.~\ref{fig:colormap}. These simulation results demonstrate the robustness and effectiveness of our algorithm.

\begin{figure*}[t]
  \centering
  \begin{overpic}[width=0.97\linewidth]{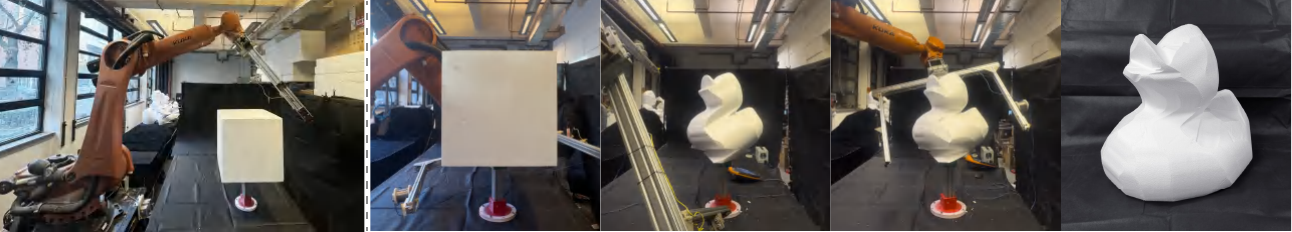}
    \put(30,1){\small  \textcolor{white}{\textbf{Init Box}}}
    \put(48,1){\small  \textcolor{white}{\textbf{4 cuts}}}
    \put(66,1){\small  \textcolor{white}{\textbf{8 cuts}}}
    \put(83,1){\small  \textcolor{white}{\textbf{Result}}}
  \end{overpic}
  \vspace{-2mm}
  \caption{
    Robotic arm product model: KUKA\textsuperscript{\textregistered} Quantec KR120 R2700 extra HA Robot (left). The fabrication process of the duck model (right).
    }
  \label{fig:Hardware}
\end{figure*}

\begin{figure}[t]
  \centering
  \begin{overpic}[width=1.0\linewidth]{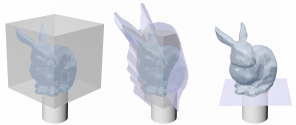}
    {
     \put(17,-3.5){\small \textbf{(a)}}
      \put(54,-3.5){\small \textbf{(b)}}
 \put(84.5,-3.5){\small \textbf{(c)}}
    }
  \end{overpic}
  \vspace{-3.5mm}
  \caption{
  (a) The material is placed on a cylindrical workbench during fabrication.
  (b) We ensure the ruled surfaces do not collide with the workbench to satisfy fabrication constraints.
  (c) The last cut path is the bottom plane of the input shape, which is used to cut off our result from the material.
  }
  \label{fig:frab-constrain}
\end{figure}

\paragraph{Hardware setup}
The rough machining machine we used in our experiments is a KUKA\textsuperscript{\textregistered} Quantec KR120 R2700 Robot extra HA, which has a 6-degree-of-freedom arm with a working range of 2696mm.
We place a metal prism in front of it as the workbench.
A polystyrene foam cube with an edge length of 600mm is put on the prism as the material to be cut (Fig.~\ref{fig:Hardware} (left)).

%

\paragraph{Fabrication constraints}
To make our algorithm feasible, we scale the size of $\mathcal{B}$ so that the hot wire is long enough to ensure the robot arm will not collide with the material.
A plane $\mathcal{P}$ is assigned as the bottom plane of the input shape.
%
We also connect the bottom of the material $\mathcal{B}$ with the cylindrical workbench $\mathcal{W}$ to prevent the robot arm from colliding with the platform while it cuts horizontally (see Fig.~\ref{fig:frab-constrain} (a)).
During our 2D approximation step, we compute $\mathbf{g}$ from $\mathcal{M}$ and $\mathcal{W}$ to ensure the generated ruled surfaces do not collide with $\mathcal{W}$ also  (see Fig.~\ref{fig:frab-constrain} (b)).
We specify the last cutting path is the plane $\mathcal{P}$ to cut off our result from the material  (see Fig.~\ref{fig:frab-constrain} (c)).
Fig.~\ref{fig:Hardware} (right) shows an example of our physical fabrication process.


\paragraph{Analysis}
We summarize the statistics of our results in the Table~\ref{table:stat0}.
For the 22 complex models, our method finishes within 15 cuts.
The average distances from our results to the inputs are almost below $1\times 10^{-2}$.
%
%
The small number of cuts and low $d_\text{result}^\text{avg}$ prove the effectiveness of our method.
The viewpoint selection and the 2D approximation in each iteration take about 0.1 and 3.5 minutes, respectively, indicating that the 2D approximation consumes the majority of the algorithm’s total runtime.
In short, our algorithm can generate results with smaller errors and fewer cutting times in less time than manual labor, which proves the practicality and effectiveness of our algorithm.

\begin{table}[!t]
	\centering
	\resizebox{\linewidth}{!}
	{
        \renewcommand{\arraystretch}{0.20}	
        \begin{tabular}{lccccc}
        \toprule
			Models& Number of cuts & $d^\text{avg}_\text{result}$ & $d_\text{bb}$ & $t_\text{view}\text{ (min)}$& $t_\text{fitting}\text{ (min)}$\\
     \midrule
      Fig.~\ref{fig:more-test} plane& 8  & $6.28\times10^{-3}$ & 1.45   & 0.9 & 22.3\\
      Fig.~\ref{fig:more-test} venus with noise & 8 & $1.02\times10^{-2}$ & 0.9& 1.6 & 22.1\\
      Fig.~\ref{fig:more-test} pot& 11  & $6.42\times10^{-3}$ & 1.57  & 1.3 & 33.9\\
      Fig.~\ref{fig:more-test} unicorn& 12  & $8.77\times10^{-3}$ & 1.41  & 1.4 & 35.4\\
    \midrule
     Fig.~\ref{fig:Baseline} bird& 5  & $7.92\times10^{-3}$ & 1.19   & 0.6 & 14.9\\
      Fig.~\ref{fig:Baseline} octopus & 8 & $9.43\times10^{-3}$ & 1.71 & 0.9 & 23.2\\
      Fig.~\ref{fig:Baseline}  sculpture1& 7  & $7.64\times10^{-3}$ & 1.09  & 0.8 & 20.1\\
    \midrule
    Fig.~\ref{fig:colormap} elderly& 9  & $8.63\times10^{-3}$ & 1.28   & 1.0 & 27.3\\
      Fig.~\ref{fig:colormap} sculpture2 & 12 & $7.28\times10^{-2}$ & 1.18 & 1.4 & 36.5\\
      Fig.~\ref{fig:colormap}  butterfly& 6  & $4.02\times10^{-3}$ & 1.21  & 0.7 & 19.2\\
      Fig.~\ref{fig:colormap}  totoro& 5  & $9.02\times10^{-3}$ & 1.32  & 0.6 & 16.1\\
    \midrule
      Fig.~\ref{fig:gallery} bear& 8  & $9.52\times10^{-3}$ & 1.24   & 0.9 & 27.5\\
      Fig.~\ref{fig:gallery} head& 9  & $9.15\times10^{-3}$ & 1.51 & 1.1 & 27.3\\
      Fig.~\ref{fig:gallery} pig& 9  & $6.84\times10^{-3}$ & 1.48  & 1.0 & 26.8\\
      Fig.~\ref{fig:gallery} drill& 9  & $9.40\times10^{-3}$ & 1.18  & 1.0 & 24.8\\
      Fig.~\ref{fig:gallery} venus& 8 & $6.64\times10^{-3}$ & 1.21& 1.0 & 28.1\\
      Fig.~\ref{fig:gallery} dog& 8 & $8.33\times10^{-3}$ & 1.29  & 0.9 & 36.2\\
      Fig.~\ref{fig:gallery} hand& 15 & $5.41\times10^{-3}$ & 1.38 & 1.8 & 61.6\\
      Fig.~\ref{fig:gallery} bird& 11  & $5.52\times10^{-3}$ & 1.31  &1.3&30.1\\
      Fig.~\ref{fig:gallery} gril& 8  & $8.12\times10^{-3}$ & 1.49 &0.9 & 25.4\\
      Fig.~\ref{fig:gallery}  toy& 11 & $9.02\times10^{-3}$ & 1.29 & 1.2 & 32.8\\
     Fig.~\ref{fig:gallery} cad1& 8  & $5.98\times10^{-3}$ & 1.13 & 0.9 & 26.4\\
      Fig.~\ref{fig:gallery}  tortoise& 8 & $6.96\times10^{-3}$& 1.31 & 1.0 & 27.8\\
    \midrule
      Fig.~\ref{fig:Physical-fabrication} bunny & 15  & $7.97\times10^{-3}$ & 1.58  & 1.7 & 55.1\\
      Fig.~\ref{fig:Physical-fabrication}  duck& 13 & $8.70\times10^{-3}$  & 1.7  & 1.5& 38.3\\
      Fig.~\ref{fig:Physical-fabrication} cat & 8  & $5.33\times10^{-3}$ & 1.30 & 0.9 & 20.2\\
      Fig.~\ref{fig:Physical-fabrication}  artwork& 7 & $6.66\times10^{-3}$  & 1.49  & 0.8& 16.3 \\
      Fig.~\ref{fig:Physical-fabrication} moai & 7  & $6.44\times10^{-3}$ & 1.21  & 0.8 & 15.1\\
      Fig.~\ref{fig:Physical-fabrication}  frog& 9 & $6.12\times10^{-3}$  & 1.43  & 1.0& 25.4 \\
      Fig.~\ref{fig:Physical-fabrication} cad2 & 10  & $5.93\times10^{-3}$ & 1.13  & 1.2 & 35.3\\
      Fig.~\ref{fig:Physical-fabrication}  squirrel& 10 & $9.01\times10^{-3}$  & 1.48  & 1.2& 31.9 \\
      Fig.~\ref{fig:Physical-fabrication} einstein &11  & $6.56\times10^{-3}$ & 1.60  & 1.3 & 55.4\\
      Fig.~\ref{fig:Physical-fabrication}  heart& 7 & $6.14\times10^{-3}$  & 1.69  & 0.8 &23.1 \\
      \midrule
      Fig.~\ref{fig:conclusion} Sofa &10  & $1.85\times10^{-2}$ & 1.50  & 1.1 & 31.1\\
      Fig.~\ref{fig:conclusion} eight& 5 & $1.36\times10^{-2}$  & 1.13  & 0.6 &15.8 \\
      \midrule
     
    \end{tabular}
    }
    \caption{
		Overview of our 35 different input shapes.
We measure the distance from our simulation shape to the input shape to define the error. $d^\text{avg}_\text{result}$ is the average distance from our result to the input shape.
    $t_\text{view}$ and $t_\text{fitting}$ denote the time consumed by the viewpoint optimization process and the 2d fitting process, respectively. These times are measured in minutes.
  }\label{table:stat0}
\end{table}

\section{Conclusion and Discussion}\label{sec:conclusion}
We propose a novel method to generate a small set of ruled surfaces for rough machining.
Central to our algorithm is a key observation: the ruled surface generated by using the outer contour line under a specific viewpoint can remove materials effectively.
Based on this observation, we transform the ruled surfaces generation problem into iteratively performing the viewpoint selection, 2D smooth curve approximation, and ruled surface extrusion.
%
%
35 different input shapes, including 10 physical fabrication inputs, are tested on our method to prove its feasibility and practicability.

\begin{figure}[t]
  \centering
  \begin{overpic}[width=0.80\linewidth]{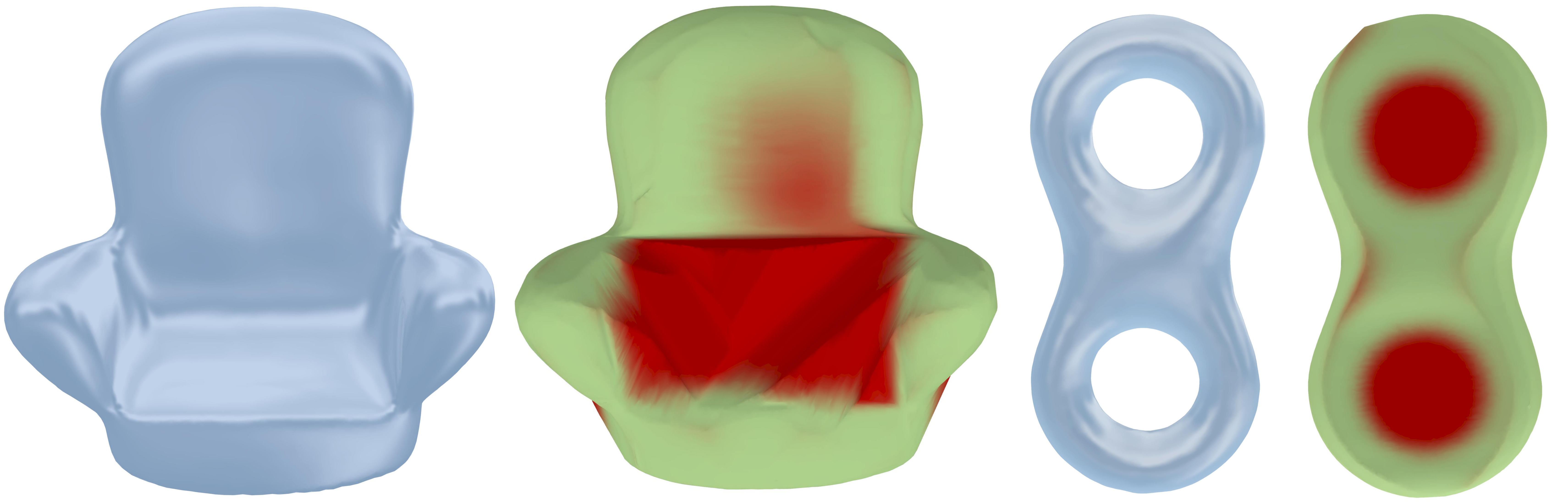}
    {
        \put(20,-3.5){\small \textbf{($1.85\times10^{-2}$, 10)}}
    \put(68,-3.5){\small \textbf{($1.36\times10^{-2}$, 5)}}
    }
  \end{overpic}
  \vspace{-0mm}
  \caption{
  Left: an example of cutting a concave shape. Right: an example of cutting a high-genus shape. The text below each result indicates
the average error and number of cuts.
  }
  \label{fig:conclusion}
\end{figure}

\paragraph{Concave and Non-zero genus shapes}
Regular hot-wire cutting equipment faces limitations when processing certain concave and non-zero genus shapes, primarily because the tool will collide with the model or prevent hot wire access to certain regions when dealing with them (see Fig.~\ref{fig:conclusion}). 
This limitation is inherently tied to the nature of hot-wire cutting technology. 
Employing curved wires rather than straight ones or segmenting the shape into multiple zero-genus components for individual cutting may partially overcome these issues, which is an intriguing direction for future research.

\paragraph{Optimal cutting path}
During machining, the end-effector tracking error does not remain constant in the cutting path.
Actually, when the cutting tool is near a singularity where the Jacobian's condition number is high, the control error increases~\cite{lynch2017modern}.
Therefore, the cutting path should be far from the singularity to decrease the fabrication error.
In the future, it would be worthwhile to develop an algorithm to generate the optimal cutting path with low control and approximation errors. 

\paragraph{Continuous cutting}
We observe that after finishing one cut, the cutting tool returns to the standby position and then starts the next cutting.
This process takes up the majority of the total time.
In addition, the order of the cutting path is determined by the user, leading to the cutting tool being more likely to travel an additional longer path when connecting two cutting paths.
We aim to study efficient cutting path planning in the future.


\paragraph{New approximation error for fine machining}
%
Our method can find a favorable configuration of the ruled surfaces to achieve a good trade-off between the number of cuts and the approximation error.
However, a suitable rough machining result for fine machining should have a uniform error distribution rather than a small average approximation error.
Proposing an approximation error that benefits fine machining to improve the overall speed of subtractive fabrication is a worthwhile research direction.



\section*{Acknowledgments}\label{sec:acknowledgments}
The authors would like to thank the anonymous reviewers for their 
constructive suggestions and comments. This work is supported by the 
National Natural Science Foundation of China (62272429, 62025207).



\begin{thebibliography}{10}
\expandafter\ifx\csname url\endcsname\relax
  \def\url#1{\texttt{#1}}\fi
\expandafter\ifx\csname urlprefix\endcsname\relax\def\urlprefix{URL }\fi
\expandafter\ifx\csname href\endcsname\relax
  \def\href#1#2{#2} \def\path#1{#1}\fi

\bibitem{yan2018multi}
C.~Yan, C.-H. Lee, X.~Li, Y.~Zhang, K.~Min, A multi-objective tool-axis optimization algorithm based on covariant field functional, Journal of Manufacturing Systems 48 (2018) 30--37.

\bibitem{LIU2021360}
D.~Liu, M.~Luo, G.~{Urbikain Pelayo}, D.~O. Trejo, D.~Zhang, Position-oriented process monitoring in milling of thin-walled parts, Journal of Manufacturing Systems 60 (2021) 360--372.

\bibitem{LI202495}
C.~Li, G.~Zhao, F.~Meng, S.~Yu, B.~Yao, H.~Liu, Multi-objective optimization of machining parameters in complete peripheral milling process with variable curvature workpieces, Journal of Manufacturing Processes 117 (2024) 95--110.

\bibitem{CHEN2018557}
Z.~Chen, Y.~Zhang, G.~Zhang, W.~Li, Modeling and reducing workpiece corner error due to wire deflection in wedm rough corner-cutting, Journal of Manufacturing Processes 36 (2018) 557--564.

\bibitem{xu2023complex}
Z.~Xu, H.~Huang, C.~Cui, X.~Liao, M.~Wu, Z.~Xue, Complex-shaped metal parts high efficiency sawing with diamond wire, International Journal of Mechanical Sciences 250 (2023) 108306.

\bibitem{zhang2025carving}
Z.~Zhang, Y.~Li, K.~Wu, H.~Zhao, X.~Liu, X.~Wang, L.~Liu, X.~Fu, Carving shapes with ruled surfaces for rough machining, Computers \& Graphics (2025) 104386.

\bibitem{van2019accessibility}
B.~van Sosin, M.~Barto{\v{n}}, G.~Elber, Accessibility for line-cutting in freeform surfaces, Comput. Aided Des. 114 (2019) 202--214.

\bibitem{duenser2020robocut}
S.~Duenser, R.~Poranne, B.~Thomaszewski, S.~Coros, Robocut: Hot-wire cutting with robot-controlled flexible rods, ACM Trans. Graph. 39~(4) (2020) 98--1.

\bibitem{lin2000ruled}
R.-S. Lin, Y.~Koren, Ruled surface machining on five-axis cnc machine tools, Journal of Manufacturing Processes 2~(1) (2000) 25--35.

\bibitem{koc2002adaptive}
B.~Koc, Y.-S. Lee, Adaptive ruled layers approximation of stl models and multiaxis machining applications for rapid prototyping, Journal of Manufacturing Systems 21~(3) (2002) 153--166.

\bibitem{sprott2008cylindrical}
K.~Sprott, B.~Ravani, Cylindrical milling of ruled surfaces, The International Journal of Advanced Manufacturing Technology 38 (2008) 649--656.

\bibitem{gong2005improved}
H.~Gong, L.-X. Cao, J.~Liu, Improved positioning of cylindrical cutter for flank milling ruled surfaces, Comput. Aided Des. 37~(12) (2005) 1205--1213.

\bibitem{li2006flank}
C.~Li, S.~Bedi, S.~Mann, Flank milling of a ruled surface with conical tools—an optimization approach, The International Journal of Advanced Manufacturing Technology 29 (2006) 1115--1124.

\bibitem{hua2018wire}
H.~Hua, T.~Jia, Wire cut of double-sided minimal surfaces, The Visual Computer 34~(6) (2018) 985--995.

\bibitem{CHU2020171}
C.-H. Chu, H.-Y. Chen, C.-H. Chang, Continuity-preserving tool path generation for minimizing machining errors in five-axis cnc flank milling of ruled surfaces, Journal of Manufacturing Systems 55 (2020) 171--178.

\bibitem{zhou2021digital}
Y.~Zhou, T.~Xing, Y.~Song, Y.~Li, X.~Zhu, G.~Li, S.~Ding, Digital-twin-driven geometric optimization of centrifugal impeller with free-form blades for five-axis flank milling, Journal of Manufacturing Systems 58 (2021) 22--35.

\bibitem{yamauchi1997frontier}
B.~Yamauchi, A frontier-based approach for autonomous exploration, in: Proceedings 1997 IEEE International Symposium on Computational Intelligence in Robotics and Automation CIRA'97.'Towards New Computational Principles for Robotics and Automation', IEEE, 1997, pp. 146--151.

\bibitem{vasquez2018tree}
J.~I. Vasquez-Gomez, L.~E. Sucar, R.~Murrieta-Cid, J.-C. Herrera-Lozada, Tree-based search of the next best view/state for three-dimensional object reconstruction, International Journal of Advanced Robotic Systems 15~(1) (2018) 1729881418754575.

\bibitem{vasquez2014view}
J.~I. Vasquez-Gomez, L.~E. Sucar, R.~Murrieta-Cid, View planning for 3d object reconstruction with a mobile manipulator robot, in: 2014 IEEE/RSJ International Conference on Intelligent Robots and Systems, IEEE, 2014, pp. 4227--4233.

\bibitem{vasquez2017view}
J.~I. Vasquez-Gomez, L.~E. Sucar, R.~Murrieta-Cid, View/state planning for three-dimensional object reconstruction under uncertainty, Autonomous Robots 41 (2017) 89--109.

\bibitem{vasquez2009view}
J.~I. V{\'a}squez-G{\'o}mez, E.~L{\"o}pez-Damian, L.~E. Sucar, View planning for 3d object reconstruction, in: 2009 IEEE/RSJ International Conference on Intelligent Robots and Systems, IEEE, 2009, pp. 4015--4020.

\bibitem{vasquez2014volumetric}
J.~I. Vasquez-Gomez, L.~E. Sucar, R.~Murrieta-Cid, E.~Lopez-Damian, Volumetric next-best-view planning for 3d object reconstruction with positioning error, International Journal of Advanced Robotic Systems 11~(10) (2014) 159.

\bibitem{potthast2014probabilistic}
C.~Potthast, G.~S. Sukhatme, A probabilistic framework for next best view estimation in a cluttered environment, Journal of Visual Communication and Image Representation 25~(1) (2014) 148--164.

\bibitem{delmerico2018comparison}
J.~Delmerico, S.~Isler, R.~Sabzevari, D.~Scaramuzza, A comparison of volumetric information gain metrics for active 3d object reconstruction, Autonomous Robots 42~(2) (2018) 197--208.

\bibitem{wu2014quality}
S.~Wu, W.~Sun, P.~Long, H.~Huang, D.~Cohen-Or, M.~Gong, O.~Deussen, B.~Chen, Quality-driven poisson-guided autoscanning, ACM Transactions on Graphics (TOG) 33~(6) (2014) 1--12.

\bibitem{maver1993occlusions}
J.~Maver, R.~Bajcsy, Occlusions as a guide for planning the next view, IEEE transactions on pattern analysis and machine intelligence 15~(5) (1993) 417--433.

\bibitem{yuan1995mechanism}
X.~Yuan, A mechanism of automatic 3d object modeling, IEEE Transactions on pattern analysis and machine intelligence 17~(3) (1995) 307--311.

\bibitem{pito1999solution}
R.~Pito, A solution to the next best view problem for automated surface acquisition, IEEE Transactions on pattern analysis and machine intelligence 21~(10) (1999) 1016--1030.

\bibitem{chen2005vision}
S.~Chen, Y.~Li, Vision sensor planning for 3-d model acquisition, IEEE Transactions on Systems, Man, and Cybernetics, Part B (Cybernetics) 35~(5) (2005) 894--904.

\bibitem{zhou2009novel}
X.~Zhou, B.~He, Y.~Li, A novel view planning method for automatic reconstruction of unknown 3-d objects based on the limit visual surface, in: 2009 Fifth International Conference on Image and Graphics, IEEE, 2009, pp. 301--306.

\bibitem{kriegel2011surface}
S.~Kriegel, T.~Bodenm{\"u}ller, M.~Suppa, G.~Hirzinger, A surface-based next-best-view approach for automated 3d model completion of unknown objects, in: 2011 IEEE International Conference on Robotics and Automation, IEEE, 2011, pp. 4869--4874.

\bibitem{mendoza2020supervised}
M.~Mendoza, J.~I. Vasquez-Gomez, H.~Taud, L.~E. Sucar, C.~Reta, Supervised learning of the next-best-view for 3d object reconstruction, Pattern Recognition Letters 133 (2020) 224--231.

\bibitem{wu20153d}
Z.~Wu, S.~Song, A.~Khosla, F.~Yu, L.~Zhang, X.~Tang, J.~Xiao, 3d shapenets: A deep representation for volumetric shapes, in: Proceedings of the IEEE conference on computer vision and pattern recognition, 2015, pp. 1912--1920.

\bibitem{zeng2020pc}
R.~Zeng, W.~Zhao, Y.-J. Liu, Pc-nbv: A point cloud based deep network for efficient next best view planning, in: 2020 IEEE/RSJ International Conference on Intelligent Robots and Systems (IROS), IEEE, 2020, pp. 7050--7057.

\bibitem{liu2024dg}
Z.~Liu, Z.~Cao, J.~Li, P.~Guan, J.~Yu, Dg-nbv: A cognitive framework for direct generation of next best view in continuous view space, IEEE Transactions on Cognitive and Developmental Systems (2024).

\bibitem{xiao2024next}
X.~Xiao, Q.~Fang, W.~Peng, Y.~Wang, L.~Hong, J.~Ye, Next best view planning via deep reinforcement leaning for 3d reconstruction of turbine blades, in: 2024 36th Chinese Control and Decision Conference (CCDC), IEEE, 2024, pp. 4022--4027.

\bibitem{li2024boundary}
L.~Li, X.~Zhang, Boundary exploration of next best view policy in 3d robotic scanning, arXiv preprint arXiv:2412.10444 (2024).

\bibitem{jin2023neu}
L.~Jin, X.~Chen, J.~R{\"u}ckin, M.~Popovi{\'c}, Neu-nbv: Next best view planning using uncertainty estimation in image-based neural rendering, in: 2023 IEEE/RSJ International Conference on Intelligent Robots and Systems (IROS), IEEE, 2023, pp. 11305--11312.

\bibitem{lee2022uncertainty}
S.~Lee, L.~Chen, J.~Wang, A.~Liniger, S.~Kumar, F.~Yu, Uncertainty guided policy for active robotic 3d reconstruction using neural radiance fields, IEEE Robotics and Automation Letters 7~(4) (2022) 12070--12077.

\bibitem{ran2023neurar}
Y.~Ran, J.~Zeng, S.~He, J.~Chen, L.~Li, Y.~Chen, G.~Lee, Q.~Ye, Neurar: Neural uncertainty for autonomous 3d reconstruction with implicit neural representations, IEEE Robotics and Automation Letters 8~(2) (2023) 1125--1132.

\bibitem{pan2022activenerf}
X.~Pan, Z.~Lai, S.~Song, G.~Huang, Activenerf: Learning where to see with uncertainty estimation, in: European Conference on Computer Vision, Springer, 2022, pp. 230--246.

\bibitem{yan2023active}
D.~Yan, J.~Liu, F.~Quan, H.~Chen, M.~Fu, Active implicit object reconstruction using uncertainty-guided next-best-view optimization, IEEE Robotics and Automation Letters (2023).

\bibitem{park2001choosing}
H.~Park, Choosing nodes and knots in closed b-spline curve interpolation to point data, Computer-Aided Design 33~(13) (2001) 967--974.

\bibitem{vassilev1996fair}
T.~I. Vassilev, Fair interpolation and approximation of b-splines by energy minimization and points insertion, Computer-Aided Design 28~(9) (1996) 753--760.

\bibitem{wang1997energy}
X.~Wang, F.~F. Cheng, B.~A. Barsky, Energy and b-spline interproximation, Computer-Aided Design 29~(7) (1997) 485--496.

\bibitem{dierckx1995curve}
P.~Dierckx, Curve and surface fitting with splines, Oxford University Press, 1995.

\bibitem{zheng2012fast}
W.~Zheng, P.~Bo, Y.~Liu, W.~Wang, Fast b-spline curve fitting by l-bfgs, Computer Aided Geometric Design 29~(7) (2012) 448--462.

\bibitem{javidrad2012accelerated}
F.~Javidrad, An accelerated simulated annealing method for b-spline curve fitting to strip-shaped scattered points, International Journal of CAD/CAM 12~(1) (2012) 9--19.

\bibitem{lin2019certified}
F.~Lin, L.-Y. Shen, C.-M. Yuan, Z.~Mi, Certified space curve fitting and trajectory planning for cnc machining with cubic b-splines, Computer-Aided Design 106 (2019) 13--29.

\bibitem{jover2022coupled}
I.~L. Jover, T.~Debarre, S.~Aziznejad, M.~Unser, Coupled splines for sparse curve fitting, IEEE Transactions on Image Processing 31 (2022) 4707--4718.

\bibitem{wang2006fitting}
W.~Wang, H.~Pottmann, Y.~Liu, Fitting b-spline curves to point clouds by curvature-based squared distance minimization, ACM Transactions on Graphics (ToG) 25~(2) (2006) 214--238.

\bibitem{pottmann2002approximation}
H.~Pottmann, S.~Leopoldseder, M.~Hofer, Approximation with active b-spline curves and surfaces, in: 10th Pacific Conference on Computer Graphics and Applications, 2002. Proceedings., IEEE, 2002, pp. 8--25.

\bibitem{liu2008revisit}
Y.~Liu, W.~Wang, A revisit to least squares orthogonal distance fitting of parametric curves and surfaces, in: Advances in Geometric Modeling and Processing: 5th International Conference, GMP 2008, Hangzhou, China, April 23-25, 2008. Proceedings 5, Springer, 2008, pp. 384--397.

\bibitem{blake2012active}
A.~Blake, M.~Isard, Active contours: the application of techniques from graphics, vision, control theory and statistics to visual tracking of shapes in motion, Springer Science \& Business Media, 2012.

\bibitem{ebrahimi2019b}
A.~Ebrahimi, G.~B. Loghmani, B-spline curve fitting by diagonal approximation bfgs methods, Iranian Journal of Science and Technology, Transactions A: Science 43 (2019) 947--958.

\bibitem{bergstrom2012fitting}
P.~Bergstr{\"o}m, I.~S{\"o}derkvist, Fitting nurbs using separable least squares techniques, International Journal of Mathematical Modelling and Numerical Optimisation 3~(4) (2012) 319--334.

\bibitem{xie2001automatic}
H.~Xie, H.~Qin, Automatic knot determination of nurbs for interactive geometric design, in: Proceedings International Conference on Shape Modeling and Applications, IEEE, 2001, pp. 267--276.

\bibitem{irshad2016outline}
M.~Irshad, S.~Khalid, M.~Z. Hussain, M.~Sarfraz, Outline capturing using rational functions with the help of genetic algorithm, Applied mathematics and computation 274 (2016) 661--678.

\bibitem{galvez2013firefly}
A.~G{\'a}lvez, A.~Iglesias, Firefly algorithm for explicit b-spline curve fitting to data points, Mathematical Problems in Engineering 2013~(1) (2013) 528215.

\bibitem{hasegawa2014bezier}
A.~Y. Hasegawa, C.~Tormena, R.~S. Parpinelli, B{\'e}zier curve parametrization using a multiobjective evolutionary algorithm., International Journal of Computer Science \& Applications 11~(2) (2014).

\bibitem{zhang2020active}
J.~Zhang, Z.~Lu, M.~Li, Active contour-based method for finger-vein image segmentation, IEEE Transactions on Instrumentation and Measurement 69~(11) (2020) 8656--8665.

\bibitem{monemian2020analysis}
M.~Monemian, H.~Rabbani, Analysis of a novel segmentation algorithm for optical coherence tomography images based on pixels intensity correlations, IEEE Transactions on Instrumentation and Measurement 70 (2020) 1--12.

\bibitem{sasmal2021detection}
P.~Sasmal, M.~K. Bhuyan, S.~Gupta, Y.~Iwahori, Detection of polyps in colonoscopic videos using saliency map-based modified particle filter, IEEE Transactions on Instrumentation and Measurement 70 (2021) 1--9.

\bibitem{yang2022active}
C.~Yang, L.~Wu, Y.~Chen, G.~Wang, G.~Weng, An active contour model based on retinex and pre-fitting reflectance for fast image segmentation, Symmetry 14~(11) (2022) 2343.

\bibitem{kass1988snakes}
M.~Kass, A.~Witkin, D.~Terzopoulos, Snakes: Active contour models, International journal of computer vision 1~(4) (1988) 321--331.

\bibitem{bayer2005laplace}
J.~D. Bayer, J.~Beaumont, A.~Krol, Laplace--dirichlet energy field specification for deformable models. an fem approach to active contour fitting, Annals of biomedical engineering 33 (2005) 1175--1186.

\bibitem{ravi2020pytorch3d}
N.~Ravi, J.~Reizenstein, D.~Novotny, T.~Gordon, W.-Y. Lo, J.~Johnson, G.~Gkioxari, Accelerating 3d deep learning with pytorch3d, arXiv:2007.08501 (2020).

\bibitem{redon2002fast}
S.~Redon, A.~Kheddar, S.~Coquillart, Fast continuous collision detection between rigid bodies, Computer Graphics Forum 21~(3) (2002) 279--287.

\bibitem{rhinoceros3d}
R.~McNeel, Associates, \href{https://www.rhino3d.com/}{Rhinoceros 3d} (2024).
\newline\urlprefix\url{https://www.rhino3d.com/}

\bibitem{liu2019soft}
S.~Liu, T.~Li, W.~Chen, H.~Li, Soft rasterizer: A differentiable renderer for image-based 3d reasoning, in: Proceedings of the IEEE/CVF international conference on computer vision, 2019, pp. 7708--7717.

\bibitem{Kingma2014AdamAM}
D.~P. Kingma, J.~Ba, \href{https://api.semanticscholar.org/CorpusID:6628106}{Adam: A method for stochastic optimization}, CoRR abs/1412.6980 (2014).
\newline\urlprefix\url{https://api.semanticscholar.org/CorpusID:6628106}

\bibitem{lynch2017modern}
K.~Lynch, Modern Robotics, Cambridge University Press, 2017.

\end{thebibliography}
\end{document}